\documentclass[reprint,amsmath,amssymb,aps]{revtex4-2}

\usepackage{graphicx}
\usepackage{dcolumn}
\usepackage{bm}
\usepackage[percent]{overpic}

\newcommand{\rme}{\text{e}}
\newcommand{\rmd}{\text{d}}

\begin{document}

\title{From an exact solution of dynamics in the vicinity of hard walls to extreme value statistics of non-Markovian processes} 

\author{Thibaut Arnoulx de Pirey}
\email{thibautdepirey@gmail.com}
\affiliation{Department of Physics, Indian Institute of Science, Bengaluru 560 012, India}


\begin{abstract}
We present an exact solution for one-dimensional overdamped dynamics near a hard wall, allowing us to connect steady-state distributions under confinement with the extreme value statistics of unconfined stochastic processes. This mapping holds regardless of the statistics of the noise driving the dynamics. We first apply this result within Brownian motion theory, deriving the noncrossing probability of a Brownian path with a specific family of curves, from which several well-known results in the field can be recovered in a unified way. We then extend the analysis to non-Markovian processes, using the mapping to a steady-state to compute the long-time noncrossing probability of a pair of run-and-tumble and Brownian particles.
\end{abstract}

\maketitle

Extreme value statistics of correlated random variables have been the subject of intense scrutiny in the physics literature (see \cite{fortin2015applications, majumdar2020extreme} for reviews on the topic). Such quantities arise in various contexts, including the characterization of the ground state of random energy landscapes \cite{bouchaud1997universality, fyodorov2008freezing, lacroix2024replica}, the study of interface height fluctuations \cite{sasamoto2010exact, majumdar2005airy}, and the investigation of first-passage times and search strategies \cite{benichou2011intermittent, tejedor2012optimizing, euijin2024}. In the context of continuous-time stochastic processes, numerous results are available for Brownian motion \cite{bray2013persistence}. In contrast, non-Markovian processes are much harder to analyze, with most results obtained perturbatively around the Brownian case. Examples include fractional Brownian motion with a Hurst exponent close to 1/2 \cite{sadhu2018generalized} and thermal active particles with small activity \cite{walter2021first}. A notable exception is the case of run-and-tumble particles, where the dynamics are driven by a noise with fixed norm and random orientation reshuffling at a constant rate \cite{solon2015active}. Extreme value statistics for these particles have been obtained in one \cite{masoliver1986first, de2021survival, PhysRevE.99.032132} and higher \cite{mori2020universal} dimensions. 

Another field that has recently received considerable attention is that of nonequilibrium steady states associated with non-Markovian dynamics in confinement. This is particularly true in the context of active matter, where particles dissipate energy in order to self-propel. The interplay between the persistence of the self-propulsion and interactions with external potentials leads to steady-state distributions that can significantly differ from their Boltzmann counterparts, for instance showing depletion at the bottom of harmonic wells for run-and-tumble particles \cite{tailleur2009sedimentation}, and a tendency to accumulate close to otherwise repulsive obstacles and walls \cite{malakar2018steady, solon2015pressure}. Interestingly, this tendency is also observed in dynamics driven by memoryless non-Gaussian noises \cite{fodor2018non}. Outside equilibrium, cases where such steady-state distributions can be obtained remain scarce, with run-and-tumble particles being a notable exception \cite{solon2015pressure}.

This work presents a universal mapping between these two key concepts in the study of stochastic processes: nonequilibrium steady states under confinement and extreme value statistics, focusing on one-dimensional systems. The mapping holds regardless of the noise statistics driving the dynamics. This result is based on an exact trajectory-wise solution for first-order dynamics near a hard-wall boundary. It therefore distinguishes itself from other duality relations, linking properties of confined and unconfined stochastic processes, previously obtained from probabilistic methods \cite{levernier2019survival, gueneau2024relating, gueneau2024bridge}.

Consider the overdamped dynamics of a particle confined in the half-line $x>0$ by a hard wall at $x = 0$ and subjected to a harmonic restoring force when $x > 0$ and a time-dependent driving force $\xi(t)$ 
\begin{eqnarray}\label{eq:EOM1}
    \dot{x} = - \mu x + \xi(t) - V'(x) \,.
\end{eqnarray}
Here $V$ is such that $V'(x>0)=0$ and $V'(x<0)= -\infty$ and accounts for the impenetrable boundary at $x=0$. More precisely, we interpret this process as the limit when $\delta t \to 0$ of the following discrete-time evolution 
\begin{equation}\label{eq:EOM_discrete}
x\left(t+\delta t\right) = \max\left(0, x(t) - \mu x(t) \delta t + \int_t^{t+\delta t}\rmd \tau \xi(\tau)\right) \,,
\end{equation}
entailing that $\dot{x} = - \mu x + \xi(t)$ whenever $x > 0$. For any piece-wise continuous driving $\xi(t)$, this is equivalent to considering Eq.~\eqref{eq:EOM1} with a steep repulsive potential $V(x)$, in the infinitely steep limit, for instance taking $V(x)=\exp(-x/\lambda)$ with $\lambda \to 0^+$. A particle at the wall therefore stays at the wall as long as $\xi(t) < 0$ and leaves it as soon as $\xi(t)$ changes sign. 

\begin{figure*}
\centering
\begin{overpic}[abs,unit=1mm,scale=0.4]{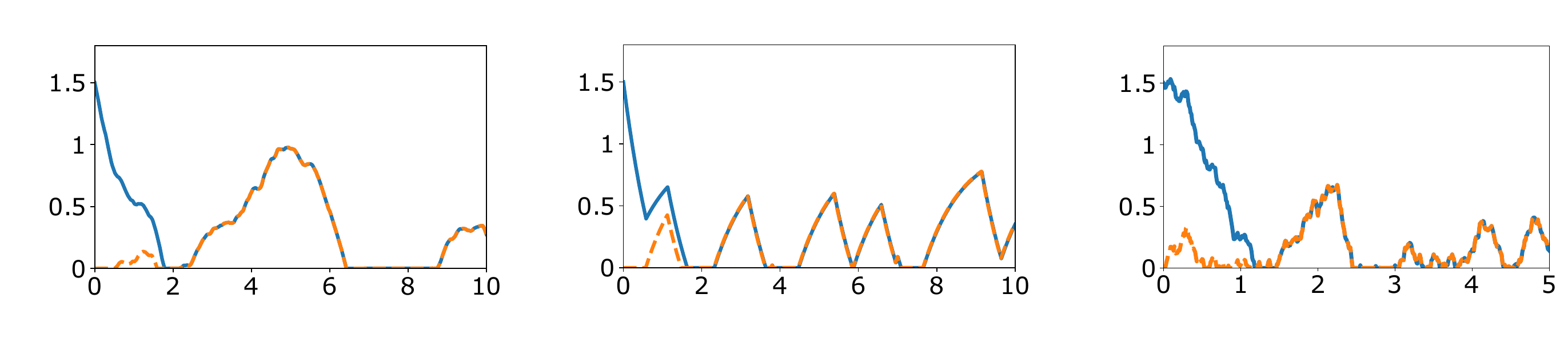}
\put(1.5,19.5){\rotatebox[origin=c]{90}{\scalebox{1.3}{$x(t)$}}}
\put(64,19.5){\rotatebox[origin=c]{90}{\scalebox{1.3}{$x(t)$}}}
\put(127,19.5){\rotatebox[origin=c]{90}{\scalebox{1.3}{$x(t)$}}}
\put(34,2){\scalebox{1.3}{$t$}}
\put(97,2){\scalebox{1.3}{$t$}}
\put(160,2){\scalebox{1.3}{$t$}}
\put(16.5,30){Ornstein-Ulhenbeck noise}
\put(82,30){run-and-tumble noise}
\put(147,30){derivative of fractional}
\put(151,27){Brownian motion}
\end{overpic}
\caption{Dynamics of a particle confined in the vicinity of a hard wall, for different types of driving processes $\xi(t)$: an Ornstein-Ulhenbeck process (Left), a telegraphic noise (Middle) and the derivative of a fractional Brownian motion with Hurst exponent $H = 0.8$ (Right). The numerical solution of the stochastic process in Eq.~\eqref{eq:EOM1} initialized at $x(0)=x_0 = 1.5$ for a specific realization of the noise $\xi(t)$ (solid, blue) is compared with the numerical evaluation of the right-hand side of Eq.~\eqref{eq:sol_direct} for the same realization of the noise $\xi(t)$ (dashed, orange). After some transient ending when the particle hits the hard wall for the first time, the two trajectories exactly match, thereby illustrating the validity of Eq.~\eqref{eq:sol_direct}. The numerical solution of Eq.~\eqref{eq:EOM1} was evaluated using the discrete-time approximation in Eq.~\eqref{eq:EOM_discrete} with $\delta t = 0.01$.}
\label{fig:cartoon}
\end{figure*}
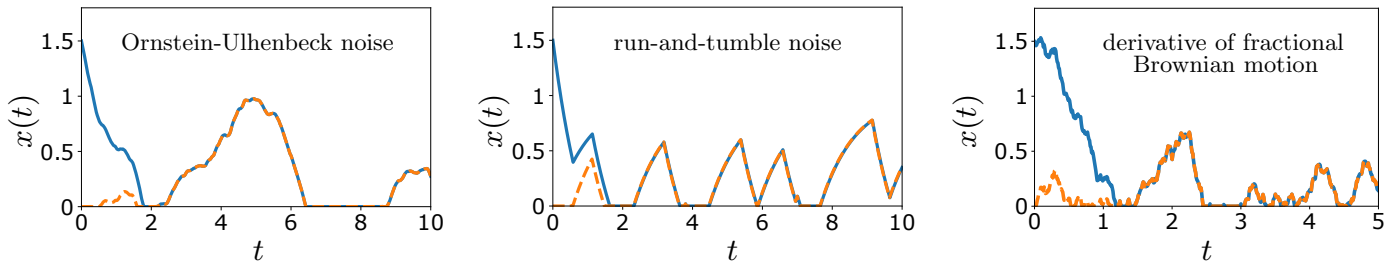

The central result of this paper rests on the fact that Eq.~\eqref{eq:EOM1}, despite being non-linear, can be integrated for any piece-wise continuous driving force $\xi(t)$. As we will prove later, if $x(0) = x_0 > 0$ and if there exists $\tau_1 >0$ such that $x(\tau_1)=0$, then for any $t>\tau_1$,
\begin{eqnarray}\label{eq:sol_direct}
    x(t) = \max_{0<t'<t}\int_0^{t'} \rmd \tau \, \rme^{-\mu \tau} \xi(t - \tau) \,.
\end{eqnarray} 
In the case where $\xi(t)$ is a time-translation invariant stochastic process, it thus follows from Eq.~\eqref{eq:sol_direct} that the steady-state cumulative distribution function associated to Eq.~\eqref{eq:EOM1} is in direct correspondence with an extreme value statistics as
\begin{eqnarray}\label{eq:mapping}
    \mathbb{P}\left[x \leq X\right] = \mathbb{P}\left[\max_{t > 0}\int_0^{t} \rmd t' \, \rme^{-\mu t'} \xi(t') \leq  X\right] \,.
\end{eqnarray}
The case of non-continuous drives, such as that of Gaussian white noise, can then be approached by studying Eq.~(\ref{eq:EOM1}) with a smoothed approximation of $\xi(t)$, for instance an Ornstein-Ulhenbeck process in the limit of vanishing correlation time. An illustration of the trajectory-wise validity of the solution in Eq.~\eqref{eq:sol_direct} is shown in Fig.~\ref{fig:cartoon}, where the function $\xi(t)$ is taken to be an Ornstein-Ulhenbeck process (Left), a telegraphic or run-and-tumble noise (Middle) and the derivative of a fractional Brownian motion (Right). 

The rest of the paper goes as follows. We first derive Eq.~\eqref{eq:sol_direct}. We then obtain from Eq.~\eqref{eq:sol_direct} a new functional identity for Brownian motion: The noncrossing probability over the time interval $[0,T]$ of a Brownian path with the curves $t \to a(1-\sqrt{1-t})$ for $a \in \mathbb{R}$ and $T \in [0,1]$, conditioned on its endpoint value. As we show next, many well-known but scattered results in the theory of Brownian motion (the joint distribution of the running maximum and the endpoint, or the probability that a Brownian curve does not cross a square root or a linear curve) can be understood as special cases of this formula. Equation \eqref{eq:mapping} is then applied in the context of non-Markovian stochastic processes by deriving the long-time noncrossing probability of a pair of run-and-tumble and Brownian particles.

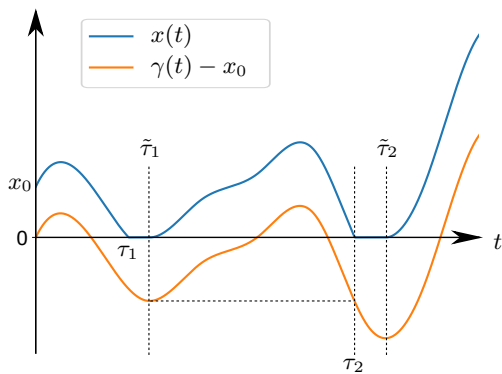
\begin{figure}
\centering
\begin{overpic}[abs,unit=1mm,scale=.4]{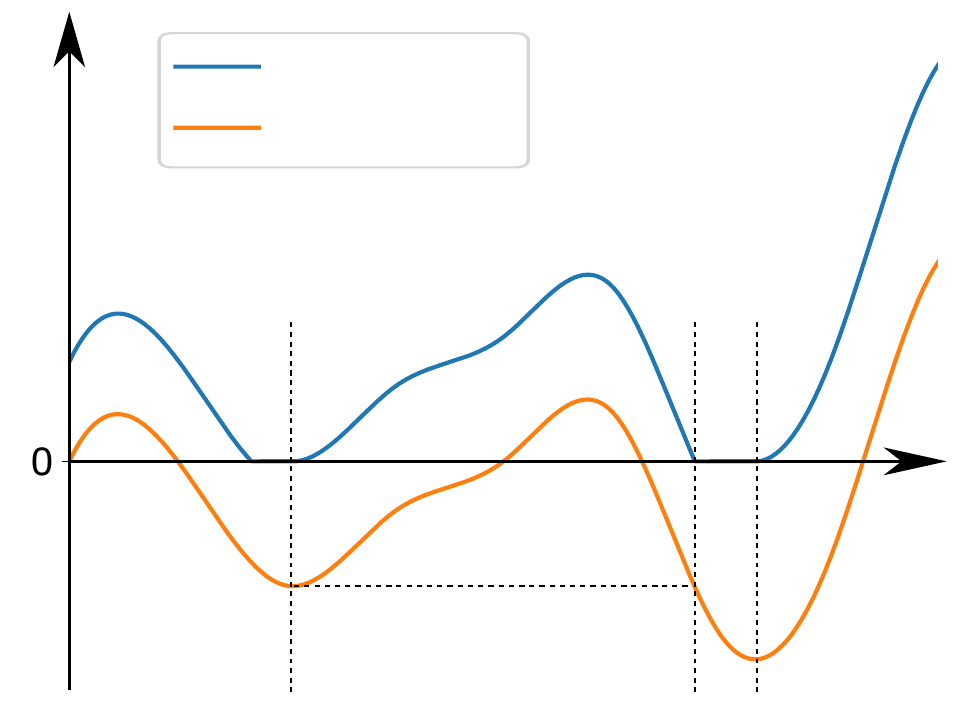}
\put(15.5,14){$\tau_1$}
\put(18.5, 27.5){$\tilde{\tau}_1$}
\put(45.5, -1){$\tau_2$}
\put(50, 27.5){$\tilde{\tau}_2$}
\put(20, 42.5){$x(t)$}
\put(1, 23){$x_0$}
\put(20, 38.3){$\gamma(t) - x_0$}
\put(65.5, 15){$t$}
\end{overpic}
\caption{Sketch of the proof of Eq.~\eqref{eq:sol_direct} for $\mu = 0$. For some chosen function $\xi(t)$, the solution $x(t)$ of the dynamics in Eq.~\eqref{eq:EOM1} is shown in blue and $\gamma(t) - x_0$ is shown in orange, with $\gamma(t)$ defined in Eq.~\eqref{eq:def_gamma}. At $\mu = 0$, whenever $x(t) > 0$, Eq.~\eqref{eq:EOM1} indicates that $\dot{x}(t) = \dot{\gamma}(t)=\xi(t)$, so that the trajectory $x(t)$, in time intervals where it does not hit the wall, is a translation of the curve $\gamma(t)$. The gap between the two curves increases each time $x(t)$ spends some time at the boundary $x=0$. The times $\{\tau_1, \tau_2,\tau_3 \dots\}$ are the increasing sequence of times at which $x(t)$ reaches $x=0$ coming from the bulk $x>0$. The times $\{\tilde{\tau}_1, \tilde{\tau}_2, \tilde{\tau}_3 \dots\}$ are the increasing sequence of times at which $x(t)$ leaves the position $x=0$. When $\dot{\gamma} > 0$, the driving force $\xi(t)$ is pushing the particle away from the wall and therefore, for any time $t>\tau_1$, the last time $\tilde{\tau}_i \leq t$ such that $x(\tilde{\tau}_i)=0$ is the argument of the minimum of $\gamma$ over the time interval $[0,t]$. Integrating Eq.~\eqref{eq:EOM1} from $\tilde{\tau}_i$ to $t$ using the condition $x(\tilde{\tau}_i)=0$ then allows to obtain Eq.~\eqref{eq:sol_direct}.}
\label{fig:fig2bis}
\end{figure}

\emph{Solution of dynamics in the vicinity of hard walls \textemdash }We start by deriving the solution in Eq.~\eqref{eq:sol_direct}.  We denote by $x_0 \geq 0$ the initial condition of Eq.~\eqref{eq:EOM1}. To solve this equation, the time axis is decomposed into intervals where the particle is away from the wall and intervals where the particle remains at $x = 0$ under the joint action of the driving force $\xi(t)$ and the confining potential $V$. Let $\tau_{i \geq 1}$ be the sequence of successive times at which the particle hits the wall coming from the bulk $x>0$ and $\tilde{\tau}_{i \geq 1}$ that of successive times at which the particle leaves the wall. By definition $\tau_i < \tilde{\tau}_i < \tau_{i+1}$. When the particle's position is $x = 0$, it remains there as long as $\xi(t) < 0$ and leaves the wall as soon as $\xi(t) > 0$. Thus $\tilde{\tau}_i = \min_{t > \tau_i}(t\vert\xi(t)>0)$. Using these definitions, and in the absence of harmonic confinement, that is for $\mu = 0$, the derivation can be conveniently pictorially summarized, see Fig.~\ref{fig:fig2bis}. In the general $\mu \neq 0$ case, we proceed by introducing 
\begin{eqnarray}\label{eq:def_gamma}
    \gamma(t) \equiv x_0 + \int_{0}^{t}\rmd\tau\,\rme^{\mu\tau}\xi(\tau)\,,
\end{eqnarray}
such that for $0 < t < \tau_1\,, \,\,\, x(t) = \gamma(t)\, \rme^{-\mu t}$
and thus $\tau_{1}={\rm min}_{t>0}(t\vert\gamma(t)=0)$. Furthermore, $x(t) = 0$ for $\tau_1 < t < \tilde{\tau}_1$ with $\tilde{\tau}_1 = \min_{t > \tau_1}(t\vert\xi(t)>0)={\rm min}_{t>\tau_{1}}(t\vert\gamma(t)\,\mbox{is a local minimum})$. For $\tilde{\tau}_1 < t < \tau_2$, Eq.~\eqref{eq:EOM1} then yields
\begin{eqnarray}
    x(t) = \rme^{-\mu t}\,\int_{\tilde{\tau}_1}^{t} \rmd \tau \,\rme^{\mu \tau}\xi(\tau) = \rme^{-\mu t}\left(\gamma(t) - \gamma(\tilde{\tau}_1) \right)\,,
\end{eqnarray}
and so $\tau_{2}={\rm min}_{t>\tilde{\tau}_1}(t\vert\gamma(t)=\gamma(\tilde{\tau}_1))$. Accordingly, $\tilde{\tau}_{2}=\min_{t > \tau_{2}}(t\vert\gamma(t)\,\text{is a local minimum}) =\min_{t>0}(t\vert\gamma(t)\,\text{is a local minimum and }\gamma(t)<\gamma(\tilde{\tau}_{1}))$. Therefore, the sequences $\tau_{i \geq 1}$ and $\tilde{\tau}_{i \geq 1}$ can be defined recursively as 
\begin{eqnarray} \label{eq:tau_sol}
    \tau_{i+1}=\min_{t>\tilde{\tau}_i}(t\vert\gamma(t)=\gamma(\tilde{\tau}_i))
\end{eqnarray}
and
\begin{equation} \label{eq:tilde_tau_sol}
    \tilde{\tau}_{i+1}=\min_{t>0}(t\vert\gamma(t)\,\mbox{is a local minimum and }\gamma(t)<\gamma(\tilde{\tau}_{i}))\,.
\end{equation}
Let now $t > \tau_1$ such that $x(t) > 0$ and let $i^* \geq 1$ such that $\tilde{\tau}_{i^*} < t < \tau_{i^*+1}$. The solution to Eq.~\eqref{eq:EOM1} is then given by
\begin{eqnarray}\label{eq:translation}
    x(t) = \rme^{-\mu t}\left(\gamma(t) - \gamma(\tilde{\tau}_{i^*}) \right) \,.
\end{eqnarray}
Furthermore, because $t < \tau_{i^*+1}$, Eqs.~(\ref{eq:tau_sol}) and (\ref{eq:tilde_tau_sol}) yield $\tilde{\tau}_{i^*}=\text{argmin}_{0<t'<t}\gamma(t')$ or equivalently $\gamma(\tilde{\tau}_{i^*}) = \text{min}_{0<t'<t}\gamma(t')$. Therefore, 
\begin{eqnarray}
   x(t) && = \rme^{-\mu t}\max_{0<t'<t}\left(\gamma(t) - \gamma(t')) \right) \,, \nonumber \\
        && = \max_{0<t'<t}\int_{t'}^t \rmd \tau \, \rme^{\mu(\tau-t)}\xi(\tau) \,,
\end{eqnarray}
from which we recover Eq.~\eqref{eq:sol_direct} after the changes of variable $\tau \to t - \tau$ and $t' \to t - t'$. Furthermore, note that if $x(t) = 0$ for $t > \tau_1$, then $\gamma(t) = \min_{0<t'<t} \gamma(t')$ so that Eq.~\eqref{eq:sol_direct} holds in general. Note also that for $t \geq \tau_1$, the system entirely loses track of its initial condition. Hence, if $x(t)$ and $y(t)$ are two realizations of Eq.~\eqref{eq:EOM1} with the same drive $\xi(t)$ but different initial conditions, $x(0) = x_0$ and $y(0)=y_0$, then $x(t) = y(t)$ for $t > {\rm max}\left(\tau_1^x, \tau_1^y\right)$ with $\tau_1^x$ (respectively $\tau_1^y$) the first time at which $x(t)=0$ (respectively $y(t)=0$). Alternatively, one can show the validity of Eq.~\eqref{eq:sol_direct} in the long-time limit from a saddle-point approximation of the solution of a well-chosen Bernoulli differential equation. This alternative derivation is presented in \cite{SM}. We believe these two proofs could serve as basis for further generalizations of our results, to other classes of stochastic processes or in higher dimension.

\begin{figure}
\centering
\begin{overpic}[abs,unit=1mm,scale=.5]{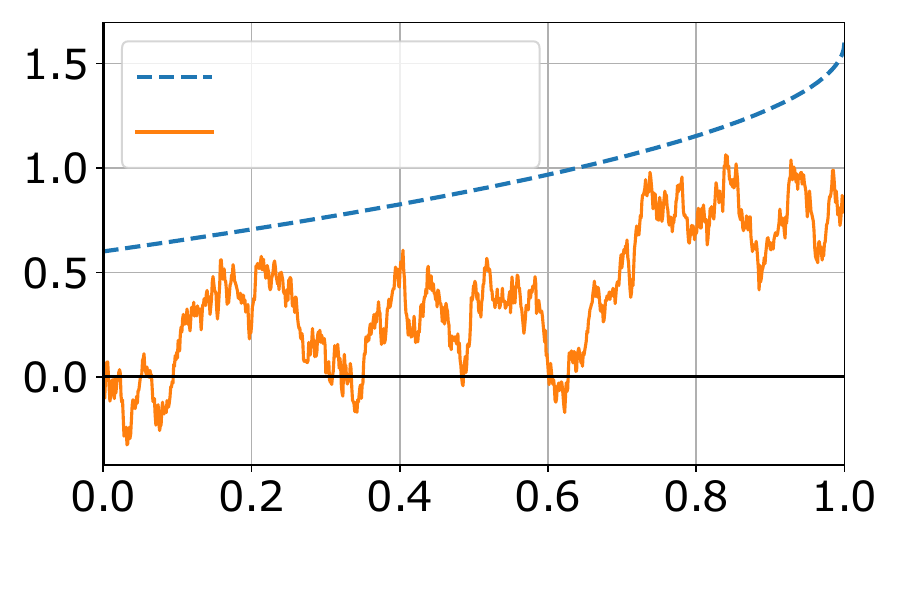}
\put(20, 38.5){Brownian motion}
\put(20,43.3){$X + 1 - \sqrt{1-t}$}
\put(39.5,3){\scalebox{1.5}{$t$}}
\end{overpic}
\caption{Illustration of a Brownian trajectory starting at the origin which remains below the curve $X + 1 - \sqrt{1-t}$ for any $0 < t < 1$.}
\label{fig:fig3}
\end{figure}

\emph{Applications to Brownian motion \textemdash }We start by applying Eqs.~(\ref{eq:sol_direct},\ref{eq:mapping}) when $\xi(t)$ is a Gaussian white noise. From now on, unless explicitly stated otherwise, we set $\mu = 1$. As we show next, Eq.~\eqref{eq:mapping} entails, as a straightforward spin-off upon a reparametrization of time, the probability distribution of the maximum of Brownian motion over a finite time interval. Consider Eq.~\eqref{eq:mapping} in the case where $\xi(t)=\sqrt{2}\eta(t)$ with $\eta(t)$ a zero-mean Gaussian white noise with correlations $\left\langle \eta(t)\eta(t') \right\rangle = \delta(t-t')$. We introduce
\begin{eqnarray}
    y(t) \equiv \sqrt{2}\int_0^{t} \rmd t' \, \rme^{-t'} \eta(t')\,,
\end{eqnarray}
for $t \geq 0$. By changing variables following $\tau' = 1 - \rme^{-2 t'}$, one gets that $y(t)$ has the same statistics as the process 
\begin{eqnarray}
    W(\tau) = \int_0^{\tau} \rmd \tau' \, \eta(\tau')\,,
\end{eqnarray}
with $\tau = 1 - \rme^{-2 t} \in [0,1]$. Here $W(\tau)$ is the standard one-dimensional Brownian motion starting at the origin. Following Eq.~\eqref{eq:mapping}, and using that the corresponding steady-state distribution of Eq.~\eqref{eq:EOM1} is given by the Boltzmann weight $P_{\rm{eq}}(x) = \Theta(x)\exp(-x^2/2)\sqrt{2/\pi}$, the distribution of the maximum of Brownian motion, for instance found in Sec. IV.B.1 in \cite{majumdar2020extreme}, is then recovered. This approach can be extended to the case where $\xi(t)$ has a non-zero mean, meaning  $\xi(t) \equiv a + \sqrt{2}\eta(t)$. In that case, 
\begin{equation}
    \int_0^{t} \rmd t' \, \rme^{-t'} \eta(t') = a \left(1-\rme^{-t}\right) +  \sqrt{2}\int_0^{t} \rmd t' \, \rme^{-t'} \eta(t')\,,
\end{equation}
and has the same statistics has 
\begin{eqnarray}
    \hat{y}(\tau) = a \left(1 - \sqrt{1-\tau}\right) + W(\tau)\,,
\end{eqnarray}
with $\tau = 1 - \rme^{-2 t} \in [0,1]$. For any $X >0$, this allows to get the probability that a Brownian particle starting at $0$ remains below the curve $t\to X - a\left(1 - \sqrt{1-\tau}\right)$, 
\begin{widetext}
\begin{equation}\label{eq:surv_curve}
    \mathbb{P}\left[\max_{0<\tau<1}\left(a \left(1 - \sqrt{1-\tau}\right) + \int_0^{\tau} \rmd \tau' \eta(\tau') \right) \leq X\right] = \frac{{\rm erf}\left(\frac{a}{\sqrt{2}}\right)+{\rm erf}\left(\frac{X-a}{\sqrt{2}}\right)}{{\rm erf} \left(\frac{a}{\sqrt{2}}\right)+1} \,,
\end{equation}
\end{widetext}
agreeing with the results of \cite{Kralchev, kahale}. An example of such noncrossing trajectory is depicted in Fig.~\ref{fig:fig3}.

\vspace{0.5cm}

To go further, note that Eq.~\eqref{eq:mapping} can be efficiently generalized at the level of two-time observables using the trajectory-wise solution Eq.~\eqref{eq:sol_direct}. In fact, we get from Eq.~\eqref{eq:sol_direct} that
\begin{eqnarray}
	x(t+\tau) && = \max \left\{\max_{0<t'<\tau} \int_0^{t'} \rmd t'' \rme^{-t''} \xi\left(t+\tau-t''\right) , \right. \nonumber \\ && \left. \max_{\tau < t' < t+\tau} \int_0^{t'}\rmd t'' \rme^{-t''} \xi\left(t+\tau-t''\right)\right\} \,.
\end{eqnarray}
Interestingly, the second term in the previous identity can be expressed in terms of $x(t)$ after the changes of variables $t'\to t'+\tau$ and $t'' \to t'' - \tau$. This leads to 
\begin{eqnarray}\label{eq:transition_proba_max}
    x(t+\tau) = \max\left\{ \max_{0<t'<\tau} z(t') \, ; \,  \rme^{-\tau}\,x(t) + z(\tau)\right\} \,,
\end{eqnarray}
with
\begin{eqnarray}
    z(t') \equiv \int_0^{t'}\rmd t'' \rme^{-t''}\xi\left(t+\tau-t''\right) \,.
\end{eqnarray}
Crucially, when $\xi(t)$ is a Gaussian white noise, $z(t')$ for $t' \in [0,\tau]$ and $x(t)$ are independent processes. Furthermore, for $\xi(t) = a + \sqrt{2}\eta(t)$, $z(t')$ has the same statistics as $a \left(1 - \sqrt{1-\tau'}\right) + W(\tau')$ with $\tau' = 1 - \rme^{-2 t'}$. Let now $P_a^w(x, \tau | x_0)$ be the transition probability associated to Eq.~\eqref{eq:EOM1}, that is the transition probability of the reflected Ornstein-Ulhenbeck process with a drift. Its expression in terms of a series expansion can be found in \cite{linetsky2005transition} and is derived in \cite{SM} for completeness. Using Eq.~\eqref{eq:transition_proba_max}, we therefore obtain the probability that a Brownian curve remains under the curve $t\to X - a\left(1 - \sqrt{1-\tau}\right)$ during a time interval $[0,T]$ and with its endpoint value being under a given threshold,
\begin{widetext}
\begin{equation}\label{eq:identity_general}
 \mathbb{P}\left[\max_{0<t<T} \left(a\left(1-\sqrt{1-t}\right) + W(t) \right) \leq X \, ; a\left(1-\sqrt{1-T}\right) + W(T) \leq X - u \right] = \int_0^X \rmd x \,  P_a^w\left(x, \left. -\ln\left(\sqrt{1-T}\right) \right| \frac{u}{\sqrt{1-T}}\right) \,,
\end{equation}
\end{widetext}
for any $0<T<1$ and $u>0$. This identity allows us to recover in a unified way, from special cases, many known results from the theory of Brownian motion. We outline these special cases here and refer the interested reader to \cite{SM} for a derivation of these quantities from Eq.~\eqref{eq:identity_general}. First, the case of a vanishing drift $a=0$, for which a simple expression for the transition probability $P_0^w$ can be obtained \cite{SM}, allows one to obtain the joint probability distribution of the maximum and the end-point of Brownian motion over some interval $[0,T]$, also present in Chapter 1.6 of \cite{harrison1985brownian}. Second, taking $u=0$ in Eq.~\eqref{eq:identity_general} generalizes to arbitrary times $0<T<1$ the identity shown in Eq.~\eqref{eq:surv_curve} and derived in \cite{Kralchev, kahale}. Third, the limit $T \to 1$ gives access to the noncrossing probability of a Brownian path with a square root curve \cite{krapivsky1996life,gautie2019non}. Lastly, the small $T \to 0$ limit, taken so that $X/\sqrt{T}$ and $a\sqrt{T}$ remain finite, yields the noncrossing probability of a Brownian path with a linear curve, see Chapter 1.8 of \cite{harrison1985brownian}.

\emph{Applications to non-Markovian processes \textemdash }We now demonstrate the use of Eq.~\eqref{eq:mapping} for the study of extreme value statistics of non-Markovian stochastic processes, focusing on processes that are diffusive over large scales. This includes, for instance, the two main models of one-dimensional active particles: the run-and-tumble and active Ornstein-Ulhenbeck \cite{martin2021statistical} ones. The results presented in this section emerge from considering the small $\mu$ limit of Eq.~\eqref{eq:mapping}. Intuitively, the left-hand side of Eq.~\eqref{eq:mapping}, up to a normalization factor, behaves as the steady-state distribution function associated with Eq.~\eqref{eq:EOM1} in the absence of harmonic confinement. Accordingly, the right-hand side of Eq.~\eqref{eq:mapping} is expected to behave as the survival probability of the dynamics $\dot{z}(t) = \xi(t)$ over a time interval $[0,T]$ with $T$ large and scaling as $T \sim \mu^{-1}$. For processes such that the dynamics of $z(t)$ is diffusive over large scales, these ideas can indeed be formalized, see \cite{SM} for details of the derivation, and we obtain
 \begin{equation}\label{eq:defS}
    \mathbb{P}\left[\max_{0< t <T}\int_0^{t} \rmd \tau \, \xi(\tau) \leq X \right] \underset{T \to \infty}{\sim} \sqrt{\frac{\beta}{T \pi}} \, \int_0^X \rmd x \, \phi(x)
\end{equation}
where $\beta^{-1} = \int_0^{+\infty} \rmd \tau \left\langle \xi(t)\xi(t+\tau)\right\rangle$ is the large scale diffusion constant of the unconstrained dynamics $z(t)$ and where $\phi(x)$ is the steady-state distribution function of non-interacting particles evolving in the vicinity of a hard wall according to 
\begin{equation}
    \dot{x} = \xi(t) - V'(x)
\end{equation}
with the boundary condition $\phi(+\infty) = 1$. This result illustrates the direct correspondence between delta-peak accumulation at the wall in the nonequilibrium steady-state described by $\phi(x)$, a prominent feature of active particles in confinement \cite{slowman2016jamming,ezhilan2015distribution,wagner2017steady}, and the non-vanishing of the survival probability when $X\to 0$.

It is interesting to note that Eq.~\eqref{eq:defS} matches the expected result over larger diffusive scales $X\sim\sqrt{T}$. Over these scales, $z(t)$ indeed behaves as Brownian motion with diffusion coefficient $\beta^{-1}$ and we get
\begin{equation}
    \lim_{T\to\infty}\mathbb{P}\left[\max_{0< t <T}\int_0^{t} \rmd \tau \, \xi(\tau) \leq X = Z\sqrt{T} \right] = \rm{erf}\left(\frac{Z\beta^{1/2}}{2}\right)\,.
\end{equation}
When $Z \ll 1$, meaning over small diffusive scales, the above equation coincides with the expression in Eq.~\eqref{eq:defS} at large subdiffusive scales $X \gg 1$, due to the boundary condition $\phi(+\infty) = 1$.

As a possible application, we note that Eq.~\eqref{eq:defS} allows us to derive the long-time noncrossing probability of a run-and-tumble particle with a Brownian one, which can be seen as the survival probability of a Brownian target initially at position $X$ chased by a run-and-tumble particle starting at the origin, see \cite{go2024active} for the related problem of the mean absorption time of a run-and-tumble particle in a thermal environment and \cite{le2019noncrossing} for that of the noncrossing probability of two run-and-tumble walkers.  The run-and-tumble dynamics at speed $v$ is described by $\dot{y} = v u(t)$ with $y(0) = 0$ and $u(t)$ a telegraphic noise switching between $\pm 1$ at rate $\tau^{-1}$, and the Brownian one by $\dot{x} = \sqrt{2D}\eta(t)$ with $x(0) = X > 0$. The noncrossing probability during time $T$ is equal to the probability that $x(t) - y(t) > 0$ for all $t \in [0,T]$, so that,
\begin{equation}\label{eq:p_noncross}
\begin{split}
& p_{\rm{non \, crossing}}(T, X) = \\ & \mathbb{P}\left[\max_{0< t <T} \left(v\int_0^{t} \rmd t' \, u(t') - \sqrt{2D}\int_0^{t} \rmd t' \, \eta(t') \right) \leq X \right]\,.
\end{split}
\end{equation} 
Using Eq.~\eqref{eq:defS}, one can therefore obtain the noncrossing probability from the steady-state distribution of particles close to a wall driven by both telegraphic and Gaussian white noises, which is derived in \cite{SM}. For an initial separation $X > 0$, we obtain
\begin{widetext}
\begin{equation}\label{eq:result_noncross}
    p_{\rm{non \, crossing}}(T, X) \underset{T \to \infty}{\sim} \sqrt{\frac{2}{T (2D + v^2\tau)\pi}} \left[X + \frac{v^ 2\tau}{2 \sqrt{v^2 + 2 D/\tau}} \left(1 - \exp\left(-\sqrt{\frac{v^2}{D^2} + \frac{2}{D\tau}}\,X\right)\right) \right] \,.
\end{equation}
\end{widetext}

\begin{figure}
\centering
\begin{overpic}[abs,unit=1mm,scale=.5]{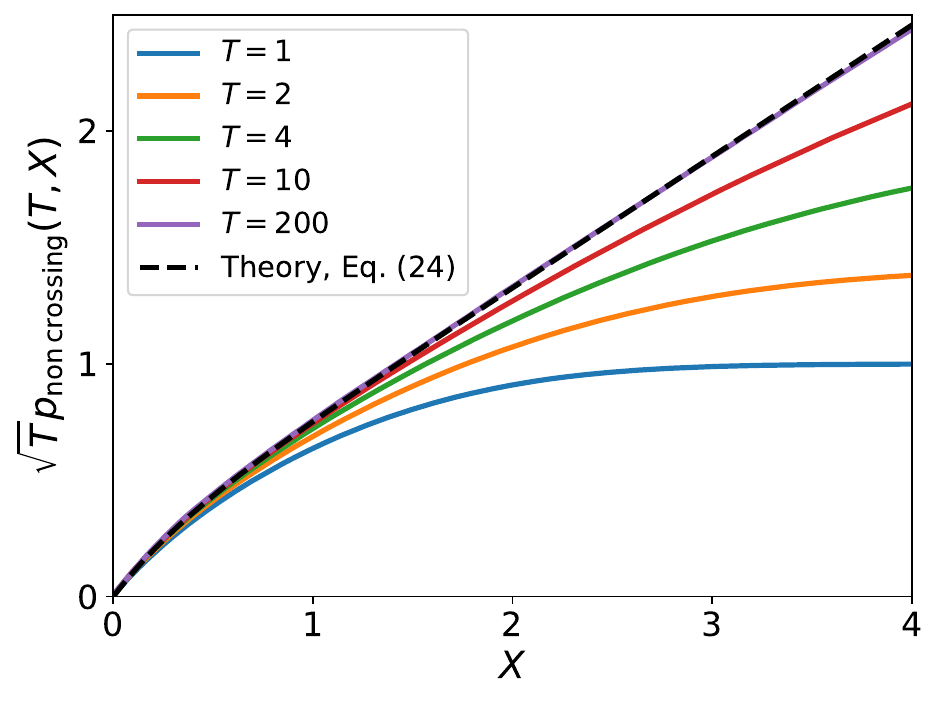}
\end{overpic}
\caption{The noncrossing probability $p_{\rm{non \, crossing}}(T, X)$ over the time interval $[0,T]$ rescaled by $\sqrt{T}$ converges at large times and finite initial separation $X$ to the analytical prediction of Eq.~\eqref{eq:result_noncross}. Here $D=1/2$, $v=1$, $\tau = 1$. To measure the noncrossing probability, $10^6$ trajectories were generated with timestep $dt = 10^{-4}$ for $t \in [0,1]$ and $dt = 10^{-3}$ for $t > 1$.}
\label{fig:fig4}
\end{figure}

\noindent We checked the validity of Eq.~\eqref{eq:result_noncross} against numerical simulations, see Fig.~\ref{fig:fig4}.

\vspace{0.5cm}
\vspace{0.5cm}

In conclusion, we have obtained an exact solution for one-dimensional overdamped dynamics near a hard wall. This solution establishes a direct connection between steady-state distributions (or transition probabilities) under confinement and extreme value statistics of unconfined stochastic processes, provided the dynamics are driven by stochastic processes with time-translation invariant statistics.

From this, we derived the noncrossing probability over $[0,T]$ of a Brownian path with the curves $t \to a(1-\sqrt{1-t})$ for $a \in \mathbb{R}$ and $T \in [0,1]$, conditioned on its endpoint value. This result offers a unified view on many classical results from Brownian motion theory, which can be recovered as special cases. Most importantly, our approach provides new insights into the extreme value statistics of non-Markovian processes, exemplified by the calculation of the long-time noncrossing probability of a pair of Brownian and run-and-tumble particles.

Future research directions include applying this mapping to study the extreme value statistics of a free underdamped Brownian particle (equivalent to an active Ornstein-Uhlenbeck process in the absence of an external potential), for which Eq.~\eqref{eq:defS} applies, and fractional Brownian motion, which lies beyond the scope of Eq.~\eqref{eq:defS} but can be approached using the more general Eq.~\eqref{eq:mapping}. It would also be interesting to explore whether the solution provided in Eq.~\eqref{eq:sol_direct} can be extended to other classes of differential equations, particularly in higher dimensions. Inspirations could be drawn from the derivations presented in this work and in \cite{SM}.

More surprisingly, we anticipate that the results presented here will be of interest in the study of collective phenomena in high-dimensional systems. Indeed, the dynamics of high-dimensional equilibrium \cite{maimbourg2016solution, agoritsas2019out, manacorda2020numerical} and active \cite{arnoulx2021active} hard-spheres, as well as that of model ecosystems with many species and random interactions \cite{arnoulx2023many}, can be framed in terms of dynamical mean-field theory equations involving one-dimensional stochastic processes with a confining boundary. The nonlinearity of these equations has hindered our understanding of such systems so far. We believe that the solution presented here will facilitate future progress.

We thank G. Bunin, G. Schehr, M. Guéneau and L. Touzo for fruitful discussions. 

\bibliography{biblio_confinement}

\begin{thebibliography}{43}%
\makeatletter
\providecommand \@ifxundefined [1]{%
 \@ifx{#1\undefined}
}%
\providecommand \@ifnum [1]{%
 \ifnum #1\expandafter \@firstoftwo
 \else \expandafter \@secondoftwo
 \fi
}%
\providecommand \@ifx [1]{%
 \ifx #1\expandafter \@firstoftwo
 \else \expandafter \@secondoftwo
 \fi
}%
\providecommand \natexlab [1]{#1}%
\providecommand \enquote  [1]{``#1''}%
\providecommand \bibnamefont  [1]{#1}%
\providecommand \bibfnamefont [1]{#1}%
\providecommand \citenamefont [1]{#1}%
\providecommand \href@noop [0]{\@secondoftwo}%
\providecommand \href [0]{\begingroup \@sanitize@url \@href}%
\providecommand \@href[1]{\@@startlink{#1}\@@href}%
\providecommand \@@href[1]{\endgroup#1\@@endlink}%
\providecommand \@sanitize@url [0]{\catcode `\\12\catcode `\$12\catcode
  `\&12\catcode `\#12\catcode `\^12\catcode `\_12\catcode `\%12\relax}%
\providecommand \@@startlink[1]{}%
\providecommand \@@endlink[0]{}%
\providecommand \url  [0]{\begingroup\@sanitize@url \@url }%
\providecommand \@url [1]{\endgroup\@href {#1}{\urlprefix }}%
\providecommand \urlprefix  [0]{URL }%
\providecommand \Eprint [0]{\href }%
\providecommand \doibase [0]{https://doi.org/}%
\providecommand \selectlanguage [0]{\@gobble}%
\providecommand \bibinfo  [0]{\@secondoftwo}%
\providecommand \bibfield  [0]{\@secondoftwo}%
\providecommand \translation [1]{[#1]}%
\providecommand \BibitemOpen [0]{}%
\providecommand \bibitemStop [0]{}%
\providecommand \bibitemNoStop [0]{.\EOS\space}%
\providecommand \EOS [0]{\spacefactor3000\relax}%
\providecommand \BibitemShut  [1]{\csname bibitem#1\endcsname}%
\let\auto@bib@innerbib\@empty
\bibitem [{\citenamefont {Fortin}\ and\ \citenamefont
  {Clusel}(2015)}]{fortin2015applications}%
  \BibitemOpen
  \bibfield  {author} {\bibinfo {author} {\bibfnamefont {J.-Y.}\ \bibnamefont
  {Fortin}}\ and\ \bibinfo {author} {\bibfnamefont {M.}~\bibnamefont
  {Clusel}},\ }\bibfield  {title} {\bibinfo {title} {Applications of extreme
  value statistics in physics},\ }\href@noop {} {\bibfield  {journal} {\bibinfo
   {journal} {Journal of Physics A: Mathematical and Theoretical}\ }\textbf
  {\bibinfo {volume} {48}},\ \bibinfo {pages} {183001} (\bibinfo {year}
  {2015})}\BibitemShut {NoStop}%
\bibitem [{\citenamefont {Majumdar}\ \emph {et~al.}(2020)\citenamefont
  {Majumdar}, \citenamefont {Pal},\ and\ \citenamefont
  {Schehr}}]{majumdar2020extreme}%
  \BibitemOpen
  \bibfield  {author} {\bibinfo {author} {\bibfnamefont {S.~N.}\ \bibnamefont
  {Majumdar}}, \bibinfo {author} {\bibfnamefont {A.}~\bibnamefont {Pal}},\ and\
  \bibinfo {author} {\bibfnamefont {G.}~\bibnamefont {Schehr}},\ }\bibfield
  {title} {\bibinfo {title} {Extreme value statistics of correlated random
  variables: a pedagogical review},\ }\href@noop {} {\bibfield  {journal}
  {\bibinfo  {journal} {Physics Reports}\ }\textbf {\bibinfo {volume} {840}},\
  \bibinfo {pages} {1} (\bibinfo {year} {2020})}\BibitemShut {NoStop}%
\bibitem [{\citenamefont {Bouchaud}\ and\ \citenamefont
  {M{\'e}zard}(1997)}]{bouchaud1997universality}%
  \BibitemOpen
  \bibfield  {author} {\bibinfo {author} {\bibfnamefont {J.-P.}\ \bibnamefont
  {Bouchaud}}\ and\ \bibinfo {author} {\bibfnamefont {M.}~\bibnamefont
  {M{\'e}zard}},\ }\bibfield  {title} {\bibinfo {title} {Universality classes
  for extreme-value statistics},\ }\href@noop {} {\bibfield  {journal}
  {\bibinfo  {journal} {Journal of Physics A: Mathematical and General}\
  }\textbf {\bibinfo {volume} {30}},\ \bibinfo {pages} {7997} (\bibinfo {year}
  {1997})}\BibitemShut {NoStop}%
\bibitem [{\citenamefont {Fyodorov}\ and\ \citenamefont
  {Bouchaud}(2008)}]{fyodorov2008freezing}%
  \BibitemOpen
  \bibfield  {author} {\bibinfo {author} {\bibfnamefont {Y.~V.}\ \bibnamefont
  {Fyodorov}}\ and\ \bibinfo {author} {\bibfnamefont {J.-P.}\ \bibnamefont
  {Bouchaud}},\ }\bibfield  {title} {\bibinfo {title} {Freezing and
  extreme-value statistics in a random energy model with logarithmically
  correlated potential},\ }\href@noop {} {\bibfield  {journal} {\bibinfo
  {journal} {Journal of Physics A: Mathematical and Theoretical}\ }\textbf
  {\bibinfo {volume} {41}},\ \bibinfo {pages} {372001} (\bibinfo {year}
  {2008})}\BibitemShut {NoStop}%
\bibitem [{\citenamefont {Lacroix-A-Chez-Toine}\ \emph
  {et~al.}(2024)\citenamefont {Lacroix-A-Chez-Toine}, \citenamefont
  {Fyodorov},\ and\ \citenamefont {Le~Doussal}}]{lacroix2024replica}%
  \BibitemOpen
  \bibfield  {author} {\bibinfo {author} {\bibfnamefont {B.}~\bibnamefont
  {Lacroix-A-Chez-Toine}}, \bibinfo {author} {\bibfnamefont {Y.~V.}\
  \bibnamefont {Fyodorov}},\ and\ \bibinfo {author} {\bibfnamefont
  {P.}~\bibnamefont {Le~Doussal}},\ }\bibfield  {title} {\bibinfo {title}
  {Replica-symmetry breaking transitions in the large deviations of the
  ground-state of a spherical spin-glass},\ }\href@noop {} {\bibfield
  {journal} {\bibinfo  {journal} {Journal of Statistical Physics}\ }\textbf
  {\bibinfo {volume} {191}},\ \bibinfo {pages} {11} (\bibinfo {year}
  {2024})}\BibitemShut {NoStop}%
\bibitem [{\citenamefont {Sasamoto}\ and\ \citenamefont
  {Spohn}(2010)}]{sasamoto2010exact}%
  \BibitemOpen
  \bibfield  {author} {\bibinfo {author} {\bibfnamefont {T.}~\bibnamefont
  {Sasamoto}}\ and\ \bibinfo {author} {\bibfnamefont {H.}~\bibnamefont
  {Spohn}},\ }\bibfield  {title} {\bibinfo {title} {Exact height distributions
  for the kpz equation with narrow wedge initial condition},\ }\href@noop {}
  {\bibfield  {journal} {\bibinfo  {journal} {Nuclear Physics B}\ }\textbf
  {\bibinfo {volume} {834}},\ \bibinfo {pages} {523} (\bibinfo {year}
  {2010})}\BibitemShut {NoStop}%
\bibitem [{\citenamefont {Majumdar}\ and\ \citenamefont
  {Comtet}(2005)}]{majumdar2005airy}%
  \BibitemOpen
  \bibfield  {author} {\bibinfo {author} {\bibfnamefont {S.~N.}\ \bibnamefont
  {Majumdar}}\ and\ \bibinfo {author} {\bibfnamefont {A.}~\bibnamefont
  {Comtet}},\ }\bibfield  {title} {\bibinfo {title} {Airy distribution
  function: from the area under a brownian excursion to the maximal height of
  fluctuating interfaces},\ }\href@noop {} {\bibfield  {journal} {\bibinfo
  {journal} {Journal of statistical physics}\ }\textbf {\bibinfo {volume}
  {119}},\ \bibinfo {pages} {777} (\bibinfo {year} {2005})}\BibitemShut
  {NoStop}%
\bibitem [{\citenamefont {B{\'e}nichou}\ \emph {et~al.}(2011)\citenamefont
  {B{\'e}nichou}, \citenamefont {Loverdo}, \citenamefont {Moreau},\ and\
  \citenamefont {Voituriez}}]{benichou2011intermittent}%
  \BibitemOpen
  \bibfield  {author} {\bibinfo {author} {\bibfnamefont {O.}~\bibnamefont
  {B{\'e}nichou}}, \bibinfo {author} {\bibfnamefont {C.}~\bibnamefont
  {Loverdo}}, \bibinfo {author} {\bibfnamefont {M.}~\bibnamefont {Moreau}},\
  and\ \bibinfo {author} {\bibfnamefont {R.}~\bibnamefont {Voituriez}},\
  }\bibfield  {title} {\bibinfo {title} {Intermittent search strategies},\
  }\href@noop {} {\bibfield  {journal} {\bibinfo  {journal} {Reviews of Modern
  Physics}\ }\textbf {\bibinfo {volume} {83}},\ \bibinfo {pages} {81} (\bibinfo
  {year} {2011})}\BibitemShut {NoStop}%
\bibitem [{\citenamefont {Tejedor}\ \emph {et~al.}(2012)\citenamefont
  {Tejedor}, \citenamefont {Voituriez},\ and\ \citenamefont
  {B{\'e}nichou}}]{tejedor2012optimizing}%
  \BibitemOpen
  \bibfield  {author} {\bibinfo {author} {\bibfnamefont {V.}~\bibnamefont
  {Tejedor}}, \bibinfo {author} {\bibfnamefont {R.}~\bibnamefont {Voituriez}},\
  and\ \bibinfo {author} {\bibfnamefont {O.}~\bibnamefont {B{\'e}nichou}},\
  }\bibfield  {title} {\bibinfo {title} {Optimizing persistent random
  searches},\ }\href@noop {} {\bibfield  {journal} {\bibinfo  {journal}
  {Physical review letters}\ }\textbf {\bibinfo {volume} {108}},\ \bibinfo
  {pages} {088103} (\bibinfo {year} {2012})}\BibitemShut {NoStop}%
\bibitem [{\citenamefont {Jeon}\ \emph {et~al.}(2024)\citenamefont {Jeon},
  \citenamefont {Go},\ and\ \citenamefont {Kim}}]{euijin2024}%
  \BibitemOpen
  \bibfield  {author} {\bibinfo {author} {\bibfnamefont {E.}~\bibnamefont
  {Jeon}}, \bibinfo {author} {\bibfnamefont {B.~G.}\ \bibnamefont {Go}},\ and\
  \bibinfo {author} {\bibfnamefont {Y.~W.}\ \bibnamefont {Kim}},\ }\bibfield
  {title} {\bibinfo {title} {Searching for a partially absorbing target by a
  run-and-tumble particle in a confined space},\ }\href
  {https://doi.org/10.1103/PhysRevE.109.014103} {\bibfield  {journal} {\bibinfo
   {journal} {Phys. Rev. E}\ }\textbf {\bibinfo {volume} {109}},\ \bibinfo
  {pages} {014103} (\bibinfo {year} {2024})}\BibitemShut {NoStop}%
\bibitem [{\citenamefont {Bray}\ \emph {et~al.}(2013)\citenamefont {Bray},
  \citenamefont {Majumdar},\ and\ \citenamefont
  {Schehr}}]{bray2013persistence}%
  \BibitemOpen
  \bibfield  {author} {\bibinfo {author} {\bibfnamefont {A.~J.}\ \bibnamefont
  {Bray}}, \bibinfo {author} {\bibfnamefont {S.~N.}\ \bibnamefont {Majumdar}},\
  and\ \bibinfo {author} {\bibfnamefont {G.}~\bibnamefont {Schehr}},\
  }\bibfield  {title} {\bibinfo {title} {Persistence and first-passage
  properties in nonequilibrium systems},\ }\href@noop {} {\bibfield  {journal}
  {\bibinfo  {journal} {Advances in Physics}\ }\textbf {\bibinfo {volume}
  {62}},\ \bibinfo {pages} {225} (\bibinfo {year} {2013})}\BibitemShut
  {NoStop}%
\bibitem [{\citenamefont {Sadhu}\ \emph {et~al.}(2018)\citenamefont {Sadhu},
  \citenamefont {Delorme},\ and\ \citenamefont {Wiese}}]{sadhu2018generalized}%
  \BibitemOpen
  \bibfield  {author} {\bibinfo {author} {\bibfnamefont {T.}~\bibnamefont
  {Sadhu}}, \bibinfo {author} {\bibfnamefont {M.}~\bibnamefont {Delorme}},\
  and\ \bibinfo {author} {\bibfnamefont {K.~J.}\ \bibnamefont {Wiese}},\
  }\bibfield  {title} {\bibinfo {title} {Generalized arcsine laws for
  fractional brownian motion},\ }\href@noop {} {\bibfield  {journal} {\bibinfo
  {journal} {Physical review letters}\ }\textbf {\bibinfo {volume} {120}},\
  \bibinfo {pages} {040603} (\bibinfo {year} {2018})}\BibitemShut {NoStop}%
\bibitem [{\citenamefont {Walter}\ \emph {et~al.}(2021)\citenamefont {Walter},
  \citenamefont {Pruessner},\ and\ \citenamefont {Salbreux}}]{walter2021first}%
  \BibitemOpen
  \bibfield  {author} {\bibinfo {author} {\bibfnamefont {B.}~\bibnamefont
  {Walter}}, \bibinfo {author} {\bibfnamefont {G.}~\bibnamefont {Pruessner}},\
  and\ \bibinfo {author} {\bibfnamefont {G.}~\bibnamefont {Salbreux}},\
  }\bibfield  {title} {\bibinfo {title} {First passage time distribution of
  active thermal particles in potentials},\ }\href@noop {} {\bibfield
  {journal} {\bibinfo  {journal} {Physical Review Research}\ }\textbf {\bibinfo
  {volume} {3}},\ \bibinfo {pages} {013075} (\bibinfo {year}
  {2021})}\BibitemShut {NoStop}%
\bibitem [{\citenamefont {Solon}\ \emph
  {et~al.}(2015{\natexlab{a}})\citenamefont {Solon}, \citenamefont {Cates},\
  and\ \citenamefont {Tailleur}}]{solon2015active}%
  \BibitemOpen
  \bibfield  {author} {\bibinfo {author} {\bibfnamefont {A.~P.}\ \bibnamefont
  {Solon}}, \bibinfo {author} {\bibfnamefont {M.~E.}\ \bibnamefont {Cates}},\
  and\ \bibinfo {author} {\bibfnamefont {J.}~\bibnamefont {Tailleur}},\
  }\bibfield  {title} {\bibinfo {title} {Active brownian particles and
  run-and-tumble particles: A comparative study},\ }\href@noop {} {\bibfield
  {journal} {\bibinfo  {journal} {The European Physical Journal Special
  Topics}\ }\textbf {\bibinfo {volume} {224}},\ \bibinfo {pages} {1231}
  (\bibinfo {year} {2015}{\natexlab{a}})}\BibitemShut {NoStop}%
\bibitem [{\citenamefont {Masoliver}\ \emph {et~al.}(1986)\citenamefont
  {Masoliver}, \citenamefont {Lindenberg},\ and\ \citenamefont
  {West}}]{masoliver1986first}%
  \BibitemOpen
  \bibfield  {author} {\bibinfo {author} {\bibfnamefont {J.}~\bibnamefont
  {Masoliver}}, \bibinfo {author} {\bibfnamefont {K.}~\bibnamefont
  {Lindenberg}},\ and\ \bibinfo {author} {\bibfnamefont {B.~J.}\ \bibnamefont
  {West}},\ }\bibfield  {title} {\bibinfo {title} {First-passage times for
  non-markovian processes: Correlated impacts on a free process},\ }\href@noop
  {} {\bibfield  {journal} {\bibinfo  {journal} {Physical Review A}\ }\textbf
  {\bibinfo {volume} {34}},\ \bibinfo {pages} {1481} (\bibinfo {year}
  {1986})}\BibitemShut {NoStop}%
\bibitem [{\citenamefont {De~Bruyne}\ \emph {et~al.}(2021)\citenamefont
  {De~Bruyne}, \citenamefont {Majumdar},\ and\ \citenamefont
  {Schehr}}]{de2021survival}%
  \BibitemOpen
  \bibfield  {author} {\bibinfo {author} {\bibfnamefont {B.}~\bibnamefont
  {De~Bruyne}}, \bibinfo {author} {\bibfnamefont {S.~N.}\ \bibnamefont
  {Majumdar}},\ and\ \bibinfo {author} {\bibfnamefont {G.}~\bibnamefont
  {Schehr}},\ }\bibfield  {title} {\bibinfo {title} {Survival probability of a
  run-and-tumble particle in the presence of a drift},\ }\href@noop {}
  {\bibfield  {journal} {\bibinfo  {journal} {Journal of Statistical Mechanics:
  Theory and Experiment}\ }\textbf {\bibinfo {volume} {2021}},\ \bibinfo
  {pages} {043211} (\bibinfo {year} {2021})}\BibitemShut {NoStop}%
\bibitem [{\citenamefont {Dhar}\ \emph {et~al.}(2019)\citenamefont {Dhar},
  \citenamefont {Kundu}, \citenamefont {Majumdar}, \citenamefont
  {Sabhapandit},\ and\ \citenamefont {Schehr}}]{PhysRevE.99.032132}%
  \BibitemOpen
  \bibfield  {author} {\bibinfo {author} {\bibfnamefont {A.}~\bibnamefont
  {Dhar}}, \bibinfo {author} {\bibfnamefont {A.}~\bibnamefont {Kundu}},
  \bibinfo {author} {\bibfnamefont {S.~N.}\ \bibnamefont {Majumdar}}, \bibinfo
  {author} {\bibfnamefont {S.}~\bibnamefont {Sabhapandit}},\ and\ \bibinfo
  {author} {\bibfnamefont {G.}~\bibnamefont {Schehr}},\ }\bibfield  {title}
  {\bibinfo {title} {Run-and-tumble particle in one-dimensional confining
  potentials: Steady-state, relaxation, and first-passage properties},\ }\href
  {https://doi.org/10.1103/PhysRevE.99.032132} {\bibfield  {journal} {\bibinfo
  {journal} {Phys. Rev. E}\ }\textbf {\bibinfo {volume} {99}},\ \bibinfo
  {pages} {032132} (\bibinfo {year} {2019})}\BibitemShut {NoStop}%
\bibitem [{\citenamefont {Mori}\ \emph {et~al.}(2020)\citenamefont {Mori},
  \citenamefont {Le~Doussal}, \citenamefont {Majumdar},\ and\ \citenamefont
  {Schehr}}]{mori2020universal}%
  \BibitemOpen
  \bibfield  {author} {\bibinfo {author} {\bibfnamefont {F.}~\bibnamefont
  {Mori}}, \bibinfo {author} {\bibfnamefont {P.}~\bibnamefont {Le~Doussal}},
  \bibinfo {author} {\bibfnamefont {S.~N.}\ \bibnamefont {Majumdar}},\ and\
  \bibinfo {author} {\bibfnamefont {G.}~\bibnamefont {Schehr}},\ }\bibfield
  {title} {\bibinfo {title} {Universal survival probability for a d-dimensional
  run-and-tumble particle},\ }\href@noop {} {\bibfield  {journal} {\bibinfo
  {journal} {Physical review letters}\ }\textbf {\bibinfo {volume} {124}},\
  \bibinfo {pages} {090603} (\bibinfo {year} {2020})}\BibitemShut {NoStop}%
\bibitem [{\citenamefont {Tailleur}\ and\ \citenamefont
  {Cates}(2009)}]{tailleur2009sedimentation}%
  \BibitemOpen
  \bibfield  {author} {\bibinfo {author} {\bibfnamefont {J.}~\bibnamefont
  {Tailleur}}\ and\ \bibinfo {author} {\bibfnamefont {M.}~\bibnamefont
  {Cates}},\ }\bibfield  {title} {\bibinfo {title} {Sedimentation, trapping,
  and rectification of dilute bacteria},\ }\href@noop {} {\bibfield  {journal}
  {\bibinfo  {journal} {Europhysics Letters}\ }\textbf {\bibinfo {volume}
  {86}},\ \bibinfo {pages} {60002} (\bibinfo {year} {2009})}\BibitemShut
  {NoStop}%
\bibitem [{\citenamefont {Malakar}\ \emph {et~al.}(2018)\citenamefont
  {Malakar}, \citenamefont {Jemseena}, \citenamefont {Kundu}, \citenamefont
  {Kumar}, \citenamefont {Sabhapandit}, \citenamefont {Majumdar}, \citenamefont
  {Redner},\ and\ \citenamefont {Dhar}}]{malakar2018steady}%
  \BibitemOpen
  \bibfield  {author} {\bibinfo {author} {\bibfnamefont {K.}~\bibnamefont
  {Malakar}}, \bibinfo {author} {\bibfnamefont {V.}~\bibnamefont {Jemseena}},
  \bibinfo {author} {\bibfnamefont {A.}~\bibnamefont {Kundu}}, \bibinfo
  {author} {\bibfnamefont {K.~V.}\ \bibnamefont {Kumar}}, \bibinfo {author}
  {\bibfnamefont {S.}~\bibnamefont {Sabhapandit}}, \bibinfo {author}
  {\bibfnamefont {S.~N.}\ \bibnamefont {Majumdar}}, \bibinfo {author}
  {\bibfnamefont {S.}~\bibnamefont {Redner}},\ and\ \bibinfo {author}
  {\bibfnamefont {A.}~\bibnamefont {Dhar}},\ }\bibfield  {title} {\bibinfo
  {title} {Steady state, relaxation and first-passage properties of a
  run-and-tumble particle in one-dimension},\ }\href@noop {} {\bibfield
  {journal} {\bibinfo  {journal} {Journal of Statistical Mechanics: Theory and
  Experiment}\ }\textbf {\bibinfo {volume} {2018}},\ \bibinfo {pages} {043215}
  (\bibinfo {year} {2018})}\BibitemShut {NoStop}%
\bibitem [{\citenamefont {Solon}\ \emph
  {et~al.}(2015{\natexlab{b}})\citenamefont {Solon}, \citenamefont {Fily},
  \citenamefont {Baskaran}, \citenamefont {Cates}, \citenamefont {Kafri},
  \citenamefont {Kardar},\ and\ \citenamefont {Tailleur}}]{solon2015pressure}%
  \BibitemOpen
  \bibfield  {author} {\bibinfo {author} {\bibfnamefont {A.~P.}\ \bibnamefont
  {Solon}}, \bibinfo {author} {\bibfnamefont {Y.}~\bibnamefont {Fily}},
  \bibinfo {author} {\bibfnamefont {A.}~\bibnamefont {Baskaran}}, \bibinfo
  {author} {\bibfnamefont {M.~E.}\ \bibnamefont {Cates}}, \bibinfo {author}
  {\bibfnamefont {Y.}~\bibnamefont {Kafri}}, \bibinfo {author} {\bibfnamefont
  {M.}~\bibnamefont {Kardar}},\ and\ \bibinfo {author} {\bibfnamefont
  {J.}~\bibnamefont {Tailleur}},\ }\bibfield  {title} {\bibinfo {title}
  {Pressure is not a state function for generic active fluids},\ }\href@noop {}
  {\bibfield  {journal} {\bibinfo  {journal} {Nature Physics}\ }\textbf
  {\bibinfo {volume} {11}},\ \bibinfo {pages} {673} (\bibinfo {year}
  {2015}{\natexlab{b}})}\BibitemShut {NoStop}%
\bibitem [{\citenamefont {Fodor}\ \emph {et~al.}(2018)\citenamefont {Fodor},
  \citenamefont {Hayakawa}, \citenamefont {Tailleur},\ and\ \citenamefont {van
  Wijland}}]{fodor2018non}%
  \BibitemOpen
  \bibfield  {author} {\bibinfo {author} {\bibfnamefont {{\'E}.}~\bibnamefont
  {Fodor}}, \bibinfo {author} {\bibfnamefont {H.}~\bibnamefont {Hayakawa}},
  \bibinfo {author} {\bibfnamefont {J.}~\bibnamefont {Tailleur}},\ and\
  \bibinfo {author} {\bibfnamefont {F.}~\bibnamefont {van Wijland}},\
  }\bibfield  {title} {\bibinfo {title} {Non-gaussian noise without memory in
  active matter},\ }\href@noop {} {\bibfield  {journal} {\bibinfo  {journal}
  {Physical Review E}\ }\textbf {\bibinfo {volume} {98}},\ \bibinfo {pages}
  {062610} (\bibinfo {year} {2018})}\BibitemShut {NoStop}%
\bibitem [{\citenamefont {Levernier}\ \emph {et~al.}(2019)\citenamefont
  {Levernier}, \citenamefont {Dolgushev}, \citenamefont {B{\'e}nichou},
  \citenamefont {Voituriez},\ and\ \citenamefont
  {Gu{\'e}rin}}]{levernier2019survival}%
  \BibitemOpen
  \bibfield  {author} {\bibinfo {author} {\bibfnamefont {N.}~\bibnamefont
  {Levernier}}, \bibinfo {author} {\bibfnamefont {M.}~\bibnamefont
  {Dolgushev}}, \bibinfo {author} {\bibfnamefont {O.}~\bibnamefont
  {B{\'e}nichou}}, \bibinfo {author} {\bibfnamefont {R.}~\bibnamefont
  {Voituriez}},\ and\ \bibinfo {author} {\bibfnamefont {T.}~\bibnamefont
  {Gu{\'e}rin}},\ }\bibfield  {title} {\bibinfo {title} {Survival probability
  of stochastic processes beyond persistence exponents},\ }\href@noop {}
  {\bibfield  {journal} {\bibinfo  {journal} {Nature communications}\ }\textbf
  {\bibinfo {volume} {10}},\ \bibinfo {pages} {2990} (\bibinfo {year}
  {2019})}\BibitemShut {NoStop}%
\bibitem [{\citenamefont {Gu{\'e}neau}\ and\ \citenamefont
  {Touzo}(2024{\natexlab{a}})}]{gueneau2024relating}%
  \BibitemOpen
  \bibfield  {author} {\bibinfo {author} {\bibfnamefont {M.}~\bibnamefont
  {Gu{\'e}neau}}\ and\ \bibinfo {author} {\bibfnamefont {L.}~\bibnamefont
  {Touzo}},\ }\bibfield  {title} {\bibinfo {title} {Relating absorbing and hard
  wall boundary conditions for a one-dimensional run-and-tumble particle},\
  }\href@noop {} {\bibfield  {journal} {\bibinfo  {journal} {Journal of Physics
  A: Mathematical and Theoretical}\ }\textbf {\bibinfo {volume} {57}},\
  \bibinfo {pages} {225005} (\bibinfo {year} {2024}{\natexlab{a}})}\BibitemShut
  {NoStop}%
\bibitem [{\citenamefont {Gu{\'e}neau}\ and\ \citenamefont
  {Touzo}(2024{\natexlab{b}})}]{gueneau2024bridge}%
  \BibitemOpen
  \bibfield  {author} {\bibinfo {author} {\bibfnamefont {M.}~\bibnamefont
  {Gu{\'e}neau}}\ and\ \bibinfo {author} {\bibfnamefont {L.}~\bibnamefont
  {Touzo}},\ }\bibfield  {title} {\bibinfo {title} {Siegmund duality for
  physicists: a bridge between spatial and first-passage properties of
  continuous-and discrete-time stochastic processes},\ }\href@noop {}
  {\bibfield  {journal} {\bibinfo  {journal} {Journal of Statistical Mechanics:
  Theory and Experiment}\ }\textbf {\bibinfo {volume} {2024}},\ \bibinfo
  {pages} {083208} (\bibinfo {year} {2024}{\natexlab{b}})}\BibitemShut
  {NoStop}%
\bibitem [{SM()}]{SM}%
  \BibitemOpen
  \href@noop {} {}\bibinfo {note} {See Supplemental Material at [URL will be
  inserted by publisher] for an alternative proof of our main result and
  demonstrations of other mathematical identities used in the main
  text.}\BibitemShut {Stop}%
\bibitem [{\citenamefont {Kralchev}(2008)}]{Kralchev}%
  \BibitemOpen
  \bibfield  {author} {\bibinfo {author} {\bibfnamefont {D.~P.}\ \bibnamefont
  {Kralchev}},\ }\bibfield  {title} {\bibinfo {title} {Levels of crossing
  probability for brownian motion},\ }\href@noop {} {\bibfield  {journal}
  {\bibinfo  {journal} {Random Operators and Stochastic Equations}\ }\textbf
  {\bibinfo {volume} {16}},\ \bibinfo {pages} {79} (\bibinfo {year}
  {2008})}\BibitemShut {NoStop}%
\bibitem [{\citenamefont {Kahale}(2008)}]{kahale}%
  \BibitemOpen
  \bibfield  {author} {\bibinfo {author} {\bibfnamefont {N.}~\bibnamefont
  {Kahale}},\ }\bibfield  {title} {\bibinfo {title} {Analytic crossing
  probabilities for certain barriers by brownian motion},\ }\href
  {http://www.jstor.org/stable/25442674} {\bibfield  {journal} {\bibinfo
  {journal} {The Annals of Applied Probability}\ }\textbf {\bibinfo {volume}
  {18}},\ \bibinfo {pages} {1424} (\bibinfo {year} {2008})}\BibitemShut
  {NoStop}%
\bibitem [{\citenamefont {Linetsky}(2005)}]{linetsky2005transition}%
  \BibitemOpen
  \bibfield  {author} {\bibinfo {author} {\bibfnamefont {V.}~\bibnamefont
  {Linetsky}},\ }\bibfield  {title} {\bibinfo {title} {On the transition
  densities for reflected diffusions},\ }\href@noop {} {\bibfield  {journal}
  {\bibinfo  {journal} {Advances in Applied Probability}\ }\textbf {\bibinfo
  {volume} {37}},\ \bibinfo {pages} {435} (\bibinfo {year} {2005})}\BibitemShut
  {NoStop}%
\bibitem [{\citenamefont {Harrison}(1985)}]{harrison1985brownian}%
  \BibitemOpen
  \bibfield  {author} {\bibinfo {author} {\bibfnamefont {J.}~\bibnamefont
  {Harrison}},\ }\bibfield  {title} {\bibinfo {title} {Brownian motion and
  stochastic flow systems},\ }\href@noop {} {\bibfield  {journal} {\bibinfo
  {journal} {Probability and Mathematical Statistics/John Wiley and Sons}\ }
  (\bibinfo {year} {1985})}\BibitemShut {NoStop}%
\bibitem [{\citenamefont {Krapivsky}\ and\ \citenamefont
  {Redner}(1996)}]{krapivsky1996life}%
  \BibitemOpen
  \bibfield  {author} {\bibinfo {author} {\bibfnamefont {P.~L.}\ \bibnamefont
  {Krapivsky}}\ and\ \bibinfo {author} {\bibfnamefont {S.}~\bibnamefont
  {Redner}},\ }\bibfield  {title} {\bibinfo {title} {Life and death in an
  expanding cage and at the edge of a receding cliff},\ }\href@noop {}
  {\bibfield  {journal} {\bibinfo  {journal} {American Journal of Physics}\
  }\textbf {\bibinfo {volume} {64}},\ \bibinfo {pages} {546} (\bibinfo {year}
  {1996})}\BibitemShut {NoStop}%
\bibitem [{\citenamefont {Gauti{\'e}}\ \emph {et~al.}(2019)\citenamefont
  {Gauti{\'e}}, \citenamefont {Le~Doussal}, \citenamefont {Majumdar},\ and\
  \citenamefont {Schehr}}]{gautie2019non}%
  \BibitemOpen
  \bibfield  {author} {\bibinfo {author} {\bibfnamefont {T.}~\bibnamefont
  {Gauti{\'e}}}, \bibinfo {author} {\bibfnamefont {P.}~\bibnamefont
  {Le~Doussal}}, \bibinfo {author} {\bibfnamefont {S.~N.}\ \bibnamefont
  {Majumdar}},\ and\ \bibinfo {author} {\bibfnamefont {G.}~\bibnamefont
  {Schehr}},\ }\bibfield  {title} {\bibinfo {title} {Non-crossing brownian
  paths and dyson brownian motion under a moving boundary},\ }\href@noop {}
  {\bibfield  {journal} {\bibinfo  {journal} {Journal of Statistical Physics}\
  }\textbf {\bibinfo {volume} {177}},\ \bibinfo {pages} {752} (\bibinfo {year}
  {2019})}\BibitemShut {NoStop}%
\bibitem [{\citenamefont {Martin}\ \emph {et~al.}(2021)\citenamefont {Martin},
  \citenamefont {O'Byrne}, \citenamefont {Cates}, \citenamefont {Fodor},
  \citenamefont {Nardini}, \citenamefont {Tailleur},\ and\ \citenamefont {van
  Wijland}}]{martin2021statistical}%
  \BibitemOpen
  \bibfield  {author} {\bibinfo {author} {\bibfnamefont {D.}~\bibnamefont
  {Martin}}, \bibinfo {author} {\bibfnamefont {J.}~\bibnamefont {O'Byrne}},
  \bibinfo {author} {\bibfnamefont {M.~E.}\ \bibnamefont {Cates}}, \bibinfo
  {author} {\bibfnamefont {E.}~\bibnamefont {Fodor}}, \bibinfo {author}
  {\bibfnamefont {C.}~\bibnamefont {Nardini}}, \bibinfo {author} {\bibfnamefont
  {J.}~\bibnamefont {Tailleur}},\ and\ \bibinfo {author} {\bibfnamefont
  {F.}~\bibnamefont {van Wijland}},\ }\bibfield  {title} {\bibinfo {title}
  {Statistical mechanics of active ornstein-uhlenbeck particles},\ }\href
  {https://doi.org/10.1103/PhysRevE.103.032607} {\bibfield  {journal} {\bibinfo
   {journal} {Phys. Rev. E}\ }\textbf {\bibinfo {volume} {103}},\ \bibinfo
  {pages} {032607} (\bibinfo {year} {2021})}\BibitemShut {NoStop}%
\bibitem [{\citenamefont {Slowman}\ \emph {et~al.}(2016)\citenamefont
  {Slowman}, \citenamefont {Evans},\ and\ \citenamefont
  {Blythe}}]{slowman2016jamming}%
  \BibitemOpen
  \bibfield  {author} {\bibinfo {author} {\bibfnamefont {A.~B.}\ \bibnamefont
  {Slowman}}, \bibinfo {author} {\bibfnamefont {M.~R.}\ \bibnamefont {Evans}},\
  and\ \bibinfo {author} {\bibfnamefont {R.~A.}\ \bibnamefont {Blythe}},\
  }\bibfield  {title} {\bibinfo {title} {Jamming and attraction of interacting
  run-and-tumble random walkers},\ }\href
  {https://doi.org/10.1103/PhysRevLett.116.218101} {\bibfield  {journal}
  {\bibinfo  {journal} {Phys. Rev. Lett.}\ }\textbf {\bibinfo {volume} {116}},\
  \bibinfo {pages} {218101} (\bibinfo {year} {2016})}\BibitemShut {NoStop}%
\bibitem [{\citenamefont {Ezhilan}\ \emph {et~al.}(2015)\citenamefont
  {Ezhilan}, \citenamefont {Alonso-Matilla},\ and\ \citenamefont
  {Saintillan}}]{ezhilan2015distribution}%
  \BibitemOpen
  \bibfield  {author} {\bibinfo {author} {\bibfnamefont {B.}~\bibnamefont
  {Ezhilan}}, \bibinfo {author} {\bibfnamefont {R.}~\bibnamefont
  {Alonso-Matilla}},\ and\ \bibinfo {author} {\bibfnamefont {D.}~\bibnamefont
  {Saintillan}},\ }\bibfield  {title} {\bibinfo {title} {On the distribution
  and swim pressure of run-and-tumble particles in confinement},\ }\href@noop
  {} {\bibfield  {journal} {\bibinfo  {journal} {Journal of Fluid Mechanics}\
  }\textbf {\bibinfo {volume} {781}},\ \bibinfo {pages} {R4} (\bibinfo {year}
  {2015})}\BibitemShut {NoStop}%
\bibitem [{\citenamefont {Wagner}\ \emph {et~al.}(2017)\citenamefont {Wagner},
  \citenamefont {Hagan},\ and\ \citenamefont {Baskaran}}]{wagner2017steady}%
  \BibitemOpen
  \bibfield  {author} {\bibinfo {author} {\bibfnamefont {C.~G.}\ \bibnamefont
  {Wagner}}, \bibinfo {author} {\bibfnamefont {M.~F.}\ \bibnamefont {Hagan}},\
  and\ \bibinfo {author} {\bibfnamefont {A.}~\bibnamefont {Baskaran}},\
  }\bibfield  {title} {\bibinfo {title} {Steady-state distributions of ideal
  active brownian particles under confinement and forcing},\ }\href@noop {}
  {\bibfield  {journal} {\bibinfo  {journal} {Journal of Statistical Mechanics:
  Theory and Experiment}\ }\textbf {\bibinfo {volume} {2017}},\ \bibinfo
  {pages} {043203} (\bibinfo {year} {2017})}\BibitemShut {NoStop}%
\bibitem [{\citenamefont {Go}\ \emph {et~al.}(2024)\citenamefont {Go},
  \citenamefont {Jeon},\ and\ \citenamefont {Kim}}]{go2024active}%
  \BibitemOpen
  \bibfield  {author} {\bibinfo {author} {\bibfnamefont {B.~G.}\ \bibnamefont
  {Go}}, \bibinfo {author} {\bibfnamefont {E.}~\bibnamefont {Jeon}},\ and\
  \bibinfo {author} {\bibfnamefont {Y.~W.}\ \bibnamefont {Kim}},\ }\bibfield
  {title} {\bibinfo {title} {Active search for a reactive target in thermal
  environments},\ }\href@noop {} {\bibfield  {journal} {\bibinfo  {journal}
  {The Journal of Chemical Physics}\ }\textbf {\bibinfo {volume} {160}}
  (\bibinfo {year} {2024})}\BibitemShut {NoStop}%
\bibitem [{\citenamefont {Le~Doussal}\ \emph {et~al.}(2019)\citenamefont
  {Le~Doussal}, \citenamefont {Majumdar},\ and\ \citenamefont
  {Schehr}}]{le2019noncrossing}%
  \BibitemOpen
  \bibfield  {author} {\bibinfo {author} {\bibfnamefont {P.}~\bibnamefont
  {Le~Doussal}}, \bibinfo {author} {\bibfnamefont {S.~N.}\ \bibnamefont
  {Majumdar}},\ and\ \bibinfo {author} {\bibfnamefont {G.}~\bibnamefont
  {Schehr}},\ }\bibfield  {title} {\bibinfo {title} {Noncrossing run-and-tumble
  particles on a line},\ }\href@noop {} {\bibfield  {journal} {\bibinfo
  {journal} {Physical Review E}\ }\textbf {\bibinfo {volume} {100}},\ \bibinfo
  {pages} {012113} (\bibinfo {year} {2019})}\BibitemShut {NoStop}%
\bibitem [{\citenamefont {Maimbourg}\ \emph {et~al.}(2016)\citenamefont
  {Maimbourg}, \citenamefont {Kurchan},\ and\ \citenamefont
  {Zamponi}}]{maimbourg2016solution}%
  \BibitemOpen
  \bibfield  {author} {\bibinfo {author} {\bibfnamefont {T.}~\bibnamefont
  {Maimbourg}}, \bibinfo {author} {\bibfnamefont {J.}~\bibnamefont {Kurchan}},\
  and\ \bibinfo {author} {\bibfnamefont {F.}~\bibnamefont {Zamponi}},\
  }\bibfield  {title} {\bibinfo {title} {Solution of the dynamics of liquids in
  the large-dimensional limit},\ }\href@noop {} {\bibfield  {journal} {\bibinfo
   {journal} {Physical review letters}\ }\textbf {\bibinfo {volume} {116}},\
  \bibinfo {pages} {015902} (\bibinfo {year} {2016})}\BibitemShut {NoStop}%
\bibitem [{\citenamefont {Agoritsas}\ \emph {et~al.}(2019)\citenamefont
  {Agoritsas}, \citenamefont {Maimbourg},\ and\ \citenamefont
  {Zamponi}}]{agoritsas2019out}%
  \BibitemOpen
  \bibfield  {author} {\bibinfo {author} {\bibfnamefont {E.}~\bibnamefont
  {Agoritsas}}, \bibinfo {author} {\bibfnamefont {T.}~\bibnamefont
  {Maimbourg}},\ and\ \bibinfo {author} {\bibfnamefont {F.}~\bibnamefont
  {Zamponi}},\ }\bibfield  {title} {\bibinfo {title} {Out-of-equilibrium
  dynamical equations of infinite-dimensional particle systems i. the isotropic
  case},\ }\href@noop {} {\bibfield  {journal} {\bibinfo  {journal} {Journal of
  Physics A: Mathematical and Theoretical}\ }\textbf {\bibinfo {volume} {52}},\
  \bibinfo {pages} {144002} (\bibinfo {year} {2019})}\BibitemShut {NoStop}%
\bibitem [{\citenamefont {Manacorda}\ \emph {et~al.}(2020)\citenamefont
  {Manacorda}, \citenamefont {Schehr},\ and\ \citenamefont
  {Zamponi}}]{manacorda2020numerical}%
  \BibitemOpen
  \bibfield  {author} {\bibinfo {author} {\bibfnamefont {A.}~\bibnamefont
  {Manacorda}}, \bibinfo {author} {\bibfnamefont {G.}~\bibnamefont {Schehr}},\
  and\ \bibinfo {author} {\bibfnamefont {F.}~\bibnamefont {Zamponi}},\
  }\bibfield  {title} {\bibinfo {title} {Numerical solution of the dynamical
  mean field theory of infinite-dimensional equilibrium liquids},\ }\href@noop
  {} {\bibfield  {journal} {\bibinfo  {journal} {The Journal of chemical
  physics}\ }\textbf {\bibinfo {volume} {152}} (\bibinfo {year}
  {2020})}\BibitemShut {NoStop}%
\bibitem [{\citenamefont {Arnoulx~de Pirey}\ \emph {et~al.}(2021)\citenamefont
  {Arnoulx~de Pirey}, \citenamefont {Manacorda}, \citenamefont {van Wijland},\
  and\ \citenamefont {Zamponi}}]{arnoulx2021active}%
  \BibitemOpen
  \bibfield  {author} {\bibinfo {author} {\bibfnamefont {T.}~\bibnamefont
  {Arnoulx~de Pirey}}, \bibinfo {author} {\bibfnamefont {A.}~\bibnamefont
  {Manacorda}}, \bibinfo {author} {\bibfnamefont {F.}~\bibnamefont {van
  Wijland}},\ and\ \bibinfo {author} {\bibfnamefont {F.}~\bibnamefont
  {Zamponi}},\ }\bibfield  {title} {\bibinfo {title} {Active matter in infinite
  dimensions: Fokker--planck equation and dynamical mean-field theory at low
  density},\ }\href@noop {} {\bibfield  {journal} {\bibinfo  {journal} {The
  Journal of Chemical Physics}\ }\textbf {\bibinfo {volume} {155}} (\bibinfo
  {year} {2021})}\BibitemShut {NoStop}%
\bibitem [{\citenamefont {Arnoulx~de Pirey}\ and\ \citenamefont
  {Bunin}(2023)}]{arnoulx2023many}%
  \BibitemOpen
  \bibfield  {author} {\bibinfo {author} {\bibfnamefont {T.}~\bibnamefont
  {Arnoulx~de Pirey}}\ and\ \bibinfo {author} {\bibfnamefont {G.}~\bibnamefont
  {Bunin}},\ }\bibfield  {title} {\bibinfo {title} {Many-species ecological
  fluctuations as a jump process from the brink of extinction},\ }\href@noop {}
  {\bibfield  {journal} {\bibinfo  {journal} {arXiv e-prints}\ ,\ \bibinfo
  {pages} {arXiv}} (\bibinfo {year} {2023})}\BibitemShut {NoStop}%
\end{thebibliography}%


\begin{thebibliography}{6}%
\makeatletter
\providecommand \@ifxundefined [1]{%
 \@ifx{#1\undefined}
}%
\providecommand \@ifnum [1]{%
 \ifnum #1\expandafter \@firstoftwo
 \else \expandafter \@secondoftwo
 \fi
}%
\providecommand \@ifx [1]{%
 \ifx #1\expandafter \@firstoftwo
 \else \expandafter \@secondoftwo
 \fi
}%
\providecommand \natexlab [1]{#1}%
\providecommand \enquote  [1]{``#1''}%
\providecommand \bibnamefont  [1]{#1}%
\providecommand \bibfnamefont [1]{#1}%
\providecommand \citenamefont [1]{#1}%
\providecommand \href@noop [0]{\@secondoftwo}%
\providecommand \href [0]{\begingroup \@sanitize@url \@href}%
\providecommand \@href[1]{\@@startlink{#1}\@@href}%
\providecommand \@@href[1]{\endgroup#1\@@endlink}%
\providecommand \@sanitize@url [0]{\catcode `\\12\catcode `\$12\catcode
  `\&12\catcode `\#12\catcode `\^12\catcode `\_12\catcode `\%12\relax}%
\providecommand \@@startlink[1]{}%
\providecommand \@@endlink[0]{}%
\providecommand \url  [0]{\begingroup\@sanitize@url \@url }%
\providecommand \@url [1]{\endgroup\@href {#1}{\urlprefix }}%
\providecommand \urlprefix  [0]{URL }%
\providecommand \Eprint [0]{\href }%
\providecommand \doibase [0]{https://doi.org/}%
\providecommand \selectlanguage [0]{\@gobble}%
\providecommand \bibinfo  [0]{\@secondoftwo}%
\providecommand \bibfield  [0]{\@secondoftwo}%
\providecommand \translation [1]{[#1]}%
\providecommand \BibitemOpen [0]{}%
\providecommand \bibitemStop [0]{}%
\providecommand \bibitemNoStop [0]{.\EOS\space}%
\providecommand \EOS [0]{\spacefactor3000\relax}%
\providecommand \BibitemShut  [1]{\csname bibitem#1\endcsname}%
\let\auto@bib@innerbib\@empty
\bibitem [{\citenamefont {Harrison}(1985)}]{harrison1985brownian}%
  \BibitemOpen
  \bibfield  {author} {\bibinfo {author} {\bibfnamefont {J.}~\bibnamefont
  {Harrison}},\ }\bibfield  {title} {\bibinfo {title} {Brownian motion and
  stochastic flow systems},\ }\href@noop {} {\bibfield  {journal} {\bibinfo
  {journal} {Probability and Mathematical Statistics/John Wiley and Sons}\ }
  (\bibinfo {year} {1985})}\BibitemShut {NoStop}%
\bibitem [{\citenamefont {Kahale}(2008)}]{kahale}%
  \BibitemOpen
  \bibfield  {author} {\bibinfo {author} {\bibfnamefont {N.}~\bibnamefont
  {Kahale}},\ }\bibfield  {title} {\bibinfo {title} {Analytic crossing
  probabilities for certain barriers by brownian motion},\ }\href
  {http://www.jstor.org/stable/25442674} {\bibfield  {journal} {\bibinfo
  {journal} {The Annals of Applied Probability}\ }\textbf {\bibinfo {volume}
  {18}},\ \bibinfo {pages} {1424} (\bibinfo {year} {2008})}\BibitemShut
  {NoStop}%
\bibitem [{\citenamefont {Kralchev}(2008)}]{Kralchev}%
  \BibitemOpen
  \bibfield  {author} {\bibinfo {author} {\bibfnamefont {D.~P.}\ \bibnamefont
  {Kralchev}},\ }\bibfield  {title} {\bibinfo {title} {Levels of crossing
  probability for brownian motion},\ }\href@noop {} {\bibfield  {journal}
  {\bibinfo  {journal} {Random Operators and Stochastic Equations}\ }\textbf
  {\bibinfo {volume} {16}},\ \bibinfo {pages} {79} (\bibinfo {year}
  {2008})}\BibitemShut {NoStop}%
\bibitem [{\citenamefont {Krapivsky}\ and\ \citenamefont
  {Redner}(1996)}]{krapivsky1996life}%
  \BibitemOpen
  \bibfield  {author} {\bibinfo {author} {\bibfnamefont {P.~L.}\ \bibnamefont
  {Krapivsky}}\ and\ \bibinfo {author} {\bibfnamefont {S.}~\bibnamefont
  {Redner}},\ }\bibfield  {title} {\bibinfo {title} {Life and death in an
  expanding cage and at the edge of a receding cliff},\ }\href@noop {}
  {\bibfield  {journal} {\bibinfo  {journal} {American Journal of Physics}\
  }\textbf {\bibinfo {volume} {64}},\ \bibinfo {pages} {546} (\bibinfo {year}
  {1996})}\BibitemShut {NoStop}%
\bibitem [{\citenamefont {Gauti{\'e}}\ \emph {et~al.}(2019)\citenamefont
  {Gauti{\'e}}, \citenamefont {Le~Doussal}, \citenamefont {Majumdar},\ and\
  \citenamefont {Schehr}}]{gautie2019non}%
  \BibitemOpen
  \bibfield  {author} {\bibinfo {author} {\bibfnamefont {T.}~\bibnamefont
  {Gauti{\'e}}}, \bibinfo {author} {\bibfnamefont {P.}~\bibnamefont
  {Le~Doussal}}, \bibinfo {author} {\bibfnamefont {S.~N.}\ \bibnamefont
  {Majumdar}},\ and\ \bibinfo {author} {\bibfnamefont {G.}~\bibnamefont
  {Schehr}},\ }\bibfield  {title} {\bibinfo {title} {Non-crossing brownian
  paths and dyson brownian motion under a moving boundary},\ }\href@noop {}
  {\bibfield  {journal} {\bibinfo  {journal} {Journal of Statistical Physics}\
  }\textbf {\bibinfo {volume} {177}},\ \bibinfo {pages} {752} (\bibinfo {year}
  {2019})}\BibitemShut {NoStop}%
\bibitem [{\citenamefont {Gu{\'e}neau}\ and\ \citenamefont
  {Touzo}(2024)}]{gueneau2024bridge}%
  \BibitemOpen
  \bibfield  {author} {\bibinfo {author} {\bibfnamefont {M.}~\bibnamefont
  {Gu{\'e}neau}}\ and\ \bibinfo {author} {\bibfnamefont {L.}~\bibnamefont
  {Touzo}},\ }\bibfield  {title} {\bibinfo {title} {Siegmund duality for
  physicists: a bridge between spatial and first-passage properties of
  continuous-and discrete-time stochastic processes},\ }\href@noop {}
  {\bibfield  {journal} {\bibinfo  {journal} {Journal of Statistical Mechanics:
  Theory and Experiment}\ }\textbf {\bibinfo {volume} {2024}},\ \bibinfo
  {pages} {083208} (\bibinfo {year} {2024})}\BibitemShut {NoStop}%
\end{thebibliography}%

\end{document}


\title{Supplemental material for ``From an exact solution of dynamics in the vicinity of hard walls to extreme value statistics of non-Markovian processes''}
\author{Thibaut Arnoulx de Pirey}
\affiliation{Department of Physics, Indian Institute of Science, Bengaluru 560 012, India}

\maketitle
\onecolumngrid

\section{Exact solution of dynamics in the vicinity of hard walls: proof from Bernoulli differential equations}

Equation~(3) of the main text can be understood as a corollary of the exact solutions of Bernoulli differential equations, as we now show. We assume that $\xi(t)$ is a continuous stochastic process with time-translation invariant statistics and we introduce the auxiliary process $\hat{\xi}(s) \equiv \xi(\ln s)$. The derivation proceeds by considering the Bernoulli differential equation,
\begin{eqnarray}\label{eq:bernoulli}
    \frac{\rmd y}{\rmd s} = - y \left(\hat{\xi}(s) + y^{n-1}\right)\,,
\end{eqnarray}
for some $n > 1$ and with initial condition $y(0) > 0$ so that $y(s) > 0$ for $s \geq 0$. This equation can be solved exactly by introducing the variable $f(s) \equiv y^{1-n}(s)$ for which the corresponding differential equation is linear
\begin{eqnarray}
\frac{\rmd f}{\rmd s} = (n-1) f(s)\hat{\xi}(s) + (n-1) \,.
\end{eqnarray} 
In terms of the variable $x(s) \equiv - s^{-1} \ln y(s)$, with $x_0 \equiv x(s_0)$ for some $s_0 > 0$, the solution to Eq.~\eqref{eq:bernoulli} thus reads
\begin{eqnarray}
    x(s) = \frac{s^{-1}}{n-1} \ln \left[ \exp\left(x_0 + (n-1)\int_{s_0}^s \rmd s' \hat{\xi}(s')\right) + (n-1)\int_{s_0}^s \rmd s' \exp\left((n-1)\int_{s'}^s \rmd s'' \hat{\xi}(s'')\right)\right] \,.
\end{eqnarray}
Reparametrizing time as $t \equiv \ln s$ yields
\begin{eqnarray} \label{eq:int_bern}
    x(t) = \frac{\rme^{-t}}{n-1} \ln \left[ \exp\left(x_0 + \rme^{t}\,(n-1)\int_0^{t - \ln s_0} \rmd \tau \, \rme^{-\tau} \xi(t-\tau)\right)\right. \nonumber \\ \left. + (n-1) \, \rme^{t} \int_0^{t - \ln s_0} \rmd \tau \, \rme^{-\tau} \exp\left(\,\rme^{t} \, (n-1)\int_0^\tau \rmd \tau' \, \rme^{-\tau'} \xi(t-\tau')\right)\right] \,.
\end{eqnarray}
Given that $\xi(t)$ has time-translation invariant correlations, and due do the factor $\rme^{t}$ in the exponential, the long-time behavior of $x(t)$ in Eq.~\eqref{eq:int_bern} can be obtained by a saddle-point calculation. The result reads
\begin{eqnarray}
    x(t) = \max_{0<\tau<t}\int_0^{\tau} \rmd \tau' \, \rme^{-\tau'} \xi(t - \tau') \,,
\end{eqnarray}
thereby recovering the right-hand side of Eq.~(3) of the main text. To recover the left-hand side, we need to prove that $x(t)$ defined in this section from the solution of Eq.~\eqref{eq:bernoulli} is indeed a solution of  Eq.~(1) of the main text, in the long-time limit. From Eq.~\eqref{eq:bernoulli}, a direct change of variable with $x(s) = - s^{-1} \ln y(s)$ and $t \equiv \ln s$ yields,
\begin{eqnarray}
    \dot{x} = - x + \xi(t) + \exp\left(-(n-1)\, \rme^t \, x(t)\right).
\end{eqnarray}
As $t\to\infty$, the last term vanishes for $x > 0$ and diverges for $x < 0$, therefore converging to the hard-wall repulsion of Eq.~(1) of the main text. This establishes Eq.~(3) of the main text for $\mu = 1$. The result can then be extended to arbitrary $\mu$ by a simple time reparametrization.

\section{Transition probability for the Ornstein-Ulhbenbeck process in confinement with a drift}
In this section, we derive the transition probability $P^w_a(x,t | x_0)$ introduced before Eq.~(19) of the main text. Consider the dynamics 
\begin{eqnarray}\label{eq:OU_wall}
    \dot{x} = -x + a + \sqrt{2}\eta(t) - V'(x)
\end{eqnarray}
confined to $x \geq 0$ by a hard wall accounted for by the term $V'(x)$ and where $\eta(t)$ is a Gaussian white noise with correlations $\left\langle \eta(t)\eta(t')\right\rangle = \delta(t-t')$. The transition probability $P^w_a(x,t | x_0)$ is the solution of
\begin{eqnarray}
         &&\partial_t P_a^w(x,t | x_0) = - \partial_x \left((a-x)P_a^w(x,t | x_0)\right) + \partial_x^2 P_a^w(x,t | x_0) \nonumber \,, \\ 
         &&\partial_x P_a^w(0,t|x_0) - a P_a^w(0,t|x_0) = 0 \nonumber \,, \\
         && P_a^w(x,0|x_0) = \delta(x-x_0) \,. 
\end{eqnarray}
Let $K(x,t|x_0) \equiv \exp\left(x^2/2 - a x\right)P_a^w(x,t | x_0)$, so that
\begin{eqnarray}
    \partial_t K = \mathcal{L}K
\end{eqnarray}
where the operator $\mathcal{L}$ reads 
\begin{eqnarray}
    & \mathcal{L}u = \rme^{\frac{x^2}{2}-ax}\left(\rme^{-\frac{x^2}{2}+ax} \, u' \right)' \, \nonumber \\
    & u'(0) = 0 \,.
\end{eqnarray}
$\mathcal{L}$ is self-adjoint for the scalar product
\begin{eqnarray}
    \langle u , v \rangle = \int_0^{+\infty} u(x) v(x) \rme^{-\frac{x^2}{2}+ax} \, \rmd x \,.
\end{eqnarray}
Let $\phi_n(x)$ for $n \geq 0$ be the normalized eigenvectors of $\mathcal{L}$ with associated eigenvalue $-\lambda_n$, that is $\mathcal{L} \phi_n(x) = - \lambda_n \phi_n(x)$ with $\phi_n'(0) = 0$. The transition probability thus reads
\begin{eqnarray}\label{eq:transition_series}
    P_a^w(x,t|x_0) = \sum_{n = 0}^{+\infty} \phi_n(x)  \rme^{-\frac{x^2}{2}+ax} \rme^{-\lambda_n t }\phi_n(x_0) \,.
\end{eqnarray}
The eigenvectors $\phi_n(x)$ satisfy
\begin{eqnarray}
    \phi_n''(x) + (a-x) \phi_n'(x) + \lambda_n \phi_n(x) = 0 \,.
\end{eqnarray}
and are thus given by
\begin{eqnarray}
    \phi_n(x) = \frac{1}{Z_n} D_{\lambda_n}(x-a)\rme^{\frac{(x-a)^2}{4}} \,,
\end{eqnarray}
with $D_{\lambda}(x)$ the parabolic cylinder function of order $\lambda$. The normalization constant reads
\begin{eqnarray}\label{eq:norm}
    Z_n = \rme^{\frac{a^2}{4}}\left[ \int_{-a}^{+\infty} \rmd x D^2_{\lambda_n}(x) \right]^{\frac{1}{2}} \,.
\end{eqnarray}
Lastly, since $\phi_n'(0) = 0$, the eigenvalues are the successive solutions of the equation 
\begin{eqnarray}\label{eq:roots}
    \lambda_n D_{\lambda_n-1}\left(-a\right) = 0 \,. 
\end{eqnarray}
In the special case where $a = 0$, meaning in the absence of drift, the series in Eq.~\eqref{eq:transition_series} can be resummed. For $a=0$, the eigenvalues are the positive even integers $\lambda_n = 2n$ for $n \geq 0$. The eigenvectors are the Hermite polynomials of degree $2n$. Up to a normalization by a factor $\sqrt{2}$, these are exactly the even eigenvectors associated to the Ornstein-Ulhenbeck dynamics in the absence of confining wall, denoted $\phi^{\rm OU}_{n}(x)$ in the following. The transition probability thus takes the form
\begin{eqnarray}\label{eq:resum}
    P_0^w(x,t | x_0) & = & 2 \sum_{p = 0}^{+\infty} \phi^{\rm OU}_{2p}(x) \phi^{\rm OU}_{2p}(x_0) \rme^{-\frac{x^2}{2}} \rme^{- 2 p t }  \,, \nonumber \\
    & = & \sum_{n = 0}^{+\infty} \phi^{\rm OU}_{n}(x) \phi^{\rm OU}_{n}(x_0) \rme^{-\frac{x^2}{2}} \rme^{- n t } + \sum_{n = 0}^{+\infty} \phi^{\rm OU}_{n}(x) \phi^{\rm OU}_{n}(-x_0) \rme^{-\frac{x^2}{2}} \rme^{- n t } \,, \nonumber \\
    & = & P^{\rm OU}(x,t|x_0) + P^{\rm OU}(x,t|-x_0) \,, \nonumber \\ 
    & = & \frac{1}{\sqrt{2\pi \left(1-\rme^{-2t}\right)}}\left[\exp\left(-\frac{\left(x-x_0\rme^{-t}\right)^2}{2\left(1-\rme^{-2t}\right)}\right)+\exp\left(-\frac{\left(x+x_0\rme^{-t}\right)^2}{2\left(1-\rme^{-2t}\right)}\right)\right] \,,
\end{eqnarray}
where $P^{\rm OU}(x,t|x_0)$ is the transition probability for the Ornstein-Ulhenbeck process in the absence of confining wall. This last result can also be obtained directly from the method of images.

\section{Special cases of equation~(19) of the main text}
We give here some mathematical details about the special cases of Eq.~(19) discussed in the main text. 

\emph{Joint distribution of the endpoint and maximum of Brownian motion \textemdash } We consider Eq.~(19) of the main text in the case where $a = 0$. This allows us to obtain the joint distribution of the endpoint and maximum of Brownian motion over the time interval $[0,T]$, which can be found in Chapter 1.6 of \cite{harrison1985brownian}. First, we get 
\begin{equation}\label{eq:identity_general}
 \mathbb{P}\left[\max_{0<t<T} W(t)  \leq X \, ; W(T) \leq X - u \right] = \int_0^X \rmd x \,  P_0^w\left(x, \left. -\ln\left(\sqrt{1-T}\right) \right| \frac{u}{\sqrt{1-T}}\right) \,.
\end{equation}
Using the expression of the transition probability obtained in Eq.~\eqref{eq:resum}, we then recover the well-known formula
\begin{eqnarray}\label{eq:joint}
    \mathbb{P}\left[\max_{0<t<T}  W(t) \leq X \, ; W(T) \leq u \right] = \frac{1}{2}\left[{\rm Erf}\left(\frac{2X-u}{\sqrt{2T}}\right)+{\rm Erf}\left(\frac{u}{\sqrt{2T}}\right)\right] \Theta(X-u) \,.
\end{eqnarray}

\emph{Generalizing Eq.~(15) of the main text to arbitrary time intervals $[0,T]$ with $0<T<1$ \textemdash }We now consider the case where $u=0$ in Eq.~(19) of the main text. When $u=0$, the condition at the endpoint of the interval becomes irrelevant and we obtain
\begin{eqnarray}\label{eq:arbitrary}
    \mathbb{P}\left[\max_{0<t<T} \left(a\left(1-\sqrt{1-t}\right) + W(t) \right) \leq X \, \right] = \int_0^X \rmd x \,  P_a^w\left(x, \left. -\ln\left(\sqrt{1-T}\right) \right| 0 \right) \,.
\end{eqnarray}
This expression generalizes to arbitrary times $T$ Eq.~(15) of the main text, which was derived in \cite{kahale,Kralchev}.

\emph{Non-crossing probability of Brownian and square-root curves \textemdash }We now demonstrate that Eq.~(19) of the main text gives access to the non-crossing probability of a Brownian path with a square root curve, see \cite{krapivsky1996life,gautie2019non}. For the sake of compactness, we first introduce the notation 
\begin{eqnarray}
    I(X,T,u,a)\equiv\int_0^X \rmd x \,  P_a^w\left(x, \left. -\ln\left(\sqrt{1-T}\right) \right| \frac{u}{\sqrt{1-T}}\right) \,.
\end{eqnarray}
From Eq.~(19) of the main text we then get
\begin{eqnarray}
    \lim_{T\to 1}I(X,T,u,a) & = & \left\langle \Theta\left[X - \max_{0<t<1} \left(a\left(1-\sqrt{1-t}\right) + W(t) \right)\right] \Theta\Big[X - \left(u + a + W(1)\right)\Big]\right\rangle \,, \nonumber\\ & = & \left\langle \Theta\left[X - \max_{0<t<1} \left(a\left(1-\sqrt{t}\right) + W(1-t) \right)\right] \Theta\Big[X - \left(u + a + W(1)\right)\Big]\right\rangle \,.
\end{eqnarray}
Thus, 
\begin{eqnarray}
    - \partial_u \lim_{T\to 1}  I(X,T,u,a) = \left\langle \Theta\left[X - \max_{0<t<1} \left(a\left(1-\sqrt{t}\right) + W(1-t) \right)\right]\delta\Big[X - \left(u + a + W(1)\right)\Big]\right\rangle  \,.
\end{eqnarray}
Therefore
\begin{eqnarray}\label{eq:square_root}
    -  \int_0^{+\infty} \rmd X \, \partial_u \left( \lim_{T\to 1} \, I(X,T,u,-a) \right) & = & \left\langle \Theta\left[u - \max_{0<t<1} \left(a\sqrt{t} + W(1-t) - W(1) \right)\right]\Theta\Big[u - a + W(1)\Big]\right\rangle \,, \nonumber\\ & = & \left\langle \Theta\left[u - \max_{0<t<1} \left(-a\sqrt{t} + W(t) \right)\right] \right\rangle \,, \nonumber\\ & = & \mathbb{P}\left[\max_{0<t<1} \left(a\sqrt{t} + W(t) \right) \leq u \, \right] \,.
\end{eqnarray}
Using the asymptotic behavior of the parabolic cylinder functions, one then gets the expansion
\begin{eqnarray}\label{eq:square_root}
    \mathbb{P}\left[\max_{0<t<1} \left(a\sqrt{t} + W(t) \right) \leq u \, \right] = \sum_{n=1}^{+\infty} \gamma_n u^{\lambda_n-1}\,,
\end{eqnarray}
where the exponents $\lambda_n$ are the strictly positive roots of Eq.~\eqref{eq:roots}, upon replacing $a\to-a$. The coefficients $\gamma_n$ are given by
\begin{eqnarray}
    \gamma_n = \lambda_n \frac{\int_a^{+\infty} \rmd x \,\, x \, D_{\lambda_n}(x)\rme^{-\frac{x^2}{4}}}{\int_a^{+\infty}\rmd x \,\, D^2_{\lambda_n}(x)}\,.
\end{eqnarray}

\emph{Non-crossing probability of Brownian and linear curves \textemdash} We finally consider Eq.~\eqref{eq:arbitrary} in the limit where $T \to 0$ with $Z \equiv X/\sqrt{T}$ and $c \equiv a\sqrt{T}/2$ remaining finite. This yields
\begin{eqnarray}\label{eq:arbitrary_limit}
    \mathbb{P}\left[\max_{0<t<1} \left(c t + W(t) \right) \leq Z \, \right] = \lim_{T\to 0}\sqrt{T}\int_0^Z \rmd z \,  P_{\frac{2c}{\sqrt{T}}}^w\left(\sqrt{T}z, \left. -\ln\left(\sqrt{1-T}\right) \right| 0 \right) \,.
\end{eqnarray}
Recall that $P_{a}^w$ is the transition probability associated to Eq.~\eqref{eq:OU_wall}. To handle the small-$T$ limit, we define the rescaled variables $s \equiv 2t/T$ and $y \equiv x/\sqrt{T}$ so that the process $y(s)$ evolves, in the $T\to 0$ limit, according to 
\begin{equation}
y'(s) = c + \eta(s) - W'(y) \,.
\end{equation}
Note that the $T \to 0$ limit effectively suppresses the harmonic confinement. We denote by $\hat{P}_{c}^w$ the corresponding transition probability. Equation \eqref{eq:arbitrary_limit} then implies that
\begin{eqnarray}\label{eq:surv_with_drift}
    \mathbb{P}\left[\max_{0<t<1} \left(c t + W(t) \right) \leq Z \, \right] = \int_0^Z \rmd y \,  \hat{P}_{c}^w\left(y, \left.1 \right| 0 \right) \,,
\end{eqnarray}  
which is a specific example of the duality relations discussed in \cite{gueneau2024bridge}.

\section{Demonstration of equation~(20) of the main text}
We now turn to the derivation of Eq.~(20) of the main text. With this goal in mind, we consider Eq.~(4) of the main text with $\mu = 1$ and a short-memory noise $\xi(t) \equiv \sqrt{\omega} \, \zeta(\omega t)$ where $\omega$ is a large parameter, which is equivalent to the small $\mu$ limit discussed in the main text. As $\omega$ is sent to infinity, $\xi(t)$ converges to a Gaussian white noise, $\lim_{\omega \to \infty} \xi(t)  = \sqrt{2 \beta^{-1}} \, \eta(t)$ with $\beta^{-1} = \int_0^{+\infty} \rmd \tau \left\langle \xi(t)\xi(t+\tau)\right\rangle$, provided the last integral exists. 

We start by relating the steady-state distribution in the left-hand side of Eq.~(4) of the main text to the steady-state distribution in the right-hand side of Eq.~(20) of the main text. We then relate the cumulative distribution of the maximum in the right-hand side of Eq.~(4) of the main text to the cumulative distribution of the maximum in the left-hand side of Eq.~(20) of the main text, thereby establishing Eq.~(20) of the main text.

To proceed with the steady-state distributions, note that to leading order as $\omega \to \infty$, and for any $x$ finite, the steady-state probability distribution of Eq.~(1) of the main text converges to the Boltzmann weight
\begin{eqnarray}\label{eq:boltzmann}
    P(x) = \sqrt{\frac{2\beta}{\pi}}\rme^{-\beta\frac{x^2}{2}} \Theta(x)\,.
\end{eqnarray}
Let us now probe the behavior of Eq.~(1) of the main text over short scales $x = z/\sqrt{\omega}$ with $z$ finite, namely in the immediate vicinity of the confining wall where the deviations to the Boltzmann weight are the most pronounced. Upon rescaling $\tau = \omega t$ the dynamics in Eq.~(1) of the main text becomes
\begin{eqnarray}
    \frac{\rmd z}{\rmd \tau} = - \frac{z}{\omega} + \zeta(\tau) - V'(z) \,.
\end{eqnarray}
Therefore, to leading order in $\omega$ at finite $z$, the steady-state distribution $P(z)$ becomes proportional to the distribution $\phi(z)$ obtained in the absence of a harmonic confinement and introduced in the main text before Eq.~(21). The proportionality constant has to be such that the distribution $P(z)$ when $z \to \infty$ matches the distribution in Eq.~\eqref{eq:boltzmann} when $x \to 0$, so that to leading order
\begin{eqnarray}\label{eq:limitPz}
    P(z) = \sqrt{\frac{2\beta}{\omega\pi}}\phi(z) \,,
\end{eqnarray}
where the $1/\sqrt{\omega}$ factor comes from the Jacobian when going from $z$ to $x$.

We now conclude by showing the correspondence between the right-hand side of Eq.~(4) of the main text and the left-hand side of Eq.~(20) of the main text. Recall that $\xi(t) = \sqrt{\omega} \, \zeta(\omega t)$ so that the right-hand side of Eq.~(4) of the main text becomes
\begin{eqnarray} \label{eq:surv1}
    \mathbb{P}\left[\max_{t > 0}\int_0^{t} \rmd \tau \, \rme^{-\tau} \xi(\tau) \leq X\right] = \mathbb{P}\left[\max_{t > 0}\int_0^{t} \rmd \tau \, \zeta(\tau) \rme^{-\frac{\tau}{\omega}} \leq Z\right]  \,,
\end{eqnarray}
with $Z = \sqrt{\omega}X$. It is clear that the right-hand side of Eq.~\eqref{eq:surv1} goes to 0 as $\omega\to\infty$, but care is required to extract the $1/\sqrt{\omega}$ leading-order behavior. Let $\epsilon > 0$ be some finite number, so that
\begin{eqnarray}\label{eq:bigstep}
    &&\mathbb{P}\left[\max_{t > 0}\int_0^{t} \rmd \tau \, \zeta(\tau) \rme^{-\frac{\tau}{\omega}} \leq Z\right] \nonumber \\ =  \,\, &&\mathbb{P}\left[\max_{0< t <\epsilon \omega}\int_0^{t} \rmd \tau \, \zeta(\tau) \rme^{-\frac{\tau}{\omega}} \leq Z \right] \mathbb{P}\left[ \left. \max_{t > \epsilon \omega}\int_0^{t} \rmd \tau \, \zeta(\tau) \rme^{-\frac{\tau}{\omega}} \leq Z  \right| \max_{0< t <\epsilon \omega}\int_0^{t} \rmd \tau \, \zeta(\tau) \rme^{-\frac{\tau}{\omega}} \leq Z \right] \,.
\end{eqnarray}
As shown next, the conditional probability in the right-hand side of the above equation admits a finite limit when $\omega\to +\infty$,
\begin{eqnarray}
    I \equiv \lim_{\omega\to\infty}\mathbb{P}\left[ \left. \max_{t > \epsilon \omega}\int_0^{t} \rmd \tau \, \zeta(\tau) \rme^{-\frac{\tau}{\omega}} \leq Z  \right| \max_{0< t <\epsilon \omega}\int_0^{t} \rmd \tau \, \zeta(\tau) \rme^{-\frac{\tau}{\omega}} \leq Z \right] \,.
\end{eqnarray}
To find the value of $I$, note that 
\begin{eqnarray}
    \max_{t > \epsilon \omega}\int_0^{t} \rmd \tau \, \zeta(\tau) \rme^{-\frac{\tau}{\omega}} & = & \int_0^{\epsilon \omega}\rmd \tau \zeta(\tau) \rme^{-\frac{\tau}{\omega}} + \max_{t > \epsilon \omega}\int_{\epsilon \omega}^{t} \rmd \tau \, \zeta(\tau) \rme^{-\frac{\tau}{\omega}}\nonumber\\ & = & \sqrt{\omega} \int_0^{\epsilon \omega} \rmd \tau \, \sqrt{\omega} \, \zeta(\omega\tau) \rme^{-\tau} + \sqrt{\omega} \, \rme^{-\epsilon}\max_{t > 0}\int_0^{t} \rmd \tau \, \sqrt{\omega} \, \zeta( \omega(\epsilon + \tau)) \rme^{-\tau} \,.
\end{eqnarray}
Therefore, 
\begin{equation}
    I = \lim_{\omega\to\infty} \mathbb{P}\left[ \left. \max_{t > 0}\int_0^{t} \rmd \tau \, \sqrt{\omega} \, \zeta( \omega(\epsilon + \tau)) \rme^{-\tau} \leq \frac{\rme^{\epsilon} Z}{\sqrt{\omega}} - \rme^{\epsilon}\int_0^{\epsilon} \rmd \tau \, \sqrt{\omega} \, \zeta(\omega\tau) \rme^{-\tau} \right| \max_{0< t <\epsilon}\int_0^{t} \rmd \tau \, \sqrt{\omega}\,\zeta(\omega\tau) \rme^{-\tau} \leq \frac{Z}{\sqrt{\omega}} \right] \,.
\end{equation}
As mentioned before, $\sqrt{\omega}\zeta(\omega t)$ converges to a Gaussian white noise with effective temperature $1/\beta$ when $\omega\to\infty$. Therefore, 
\begin{eqnarray}
    I = \lim_{a \to 0} \mathbb{P}\left[ \left. \max_{0 < t < 1} W_1(t)  \leq \rme^{\epsilon} a  - \rme^{\epsilon} \, W_2(\tau_\epsilon) \right| \max_{0< t <\tau_\epsilon}W_2(t) \leq a \right] \,,
\end{eqnarray}
where $W_1(t)$ and $W_2(t)$ are two independent Brownian motion, $a$ is a regularizing parameter and $\tau_\epsilon = 1 - \rme^{-2\epsilon}$. A reparametrization of time $\tau' = 1 - \rme^{-2\tau}$ was used to go from integrals of the Gaussian white noise with exponential kernels to Brownian motion. The value of $I$ can then be obtained from the joint distribution of the maximum of Brownian motion and its endpoint value, see Eq.~\eqref{eq:joint}. It reads
\begin{equation}\label{eq:valueI}
\begin{split}
I & = \int_{0}^{+\infty} \rmd u \, \frac{u}{\tau_\epsilon}\exp\left(-\frac{u^2}{2\tau_\epsilon}\right) {\rm Erf}\left(\frac{\rme^\epsilon u}{\sqrt{2}}\right) \,, \\ & = \frac{\rme^\epsilon}{\sqrt{\rme^{2\epsilon}+\tau^{-1}_\epsilon}} \,.
\end{split}
\end{equation}
Hence, from Eq.~\eqref{eq:bigstep} together with Eq.~\eqref{eq:valueI} and Eq.~\eqref{eq:limitPz}, we get to leading order as $\omega \to \infty$
\begin{eqnarray}
    \mathbb{P}\left[\max_{0< t <\epsilon \omega}\int_0^{t} \rmd \tau \, \zeta(\tau) \rme^{-\frac{\tau}{\omega}} \leq Z \right] \underset{\omega \to \infty}{\sim}  \rme^{-\epsilon}\sqrt{\frac{2\left(\rme^{2\epsilon}+\tau^{-1}_\epsilon\right)\beta}{\omega\pi}}\int_0^Z \rmd z \, \phi(z)
\end{eqnarray}
By denoting $T = \epsilon \omega$ and taking $\epsilon \to 0$, Eq.~(20) of the main text is then recovered.

\section{Steady-state distribution in the vicinity of a wall of run-and-tumble particles with thermal noise}\label{sec:boundary_layer}
In this section, we derive the steady-state density profile of non-interacting run-and-tumble particles subjected to thermal noise in the vicinity of a wall with a fixed nonzero density far away from it. Using Eq.~(20) of the main text, this allows us to address the problem of the long-time noncrossing probability of a Brownian particle with a run-and-tumble one and obtain Eq. (24) of the main text.
\noindent Consider the dynamics 
\begin{eqnarray}
    \dot{x} =  v u - V'(x) + \sqrt{2D}\eta(t)
\end{eqnarray}
where $u$ is a telegraphic process with rate $\tau^{-1}$ and $\eta(t)$ a Gaussian white noise. As before, $V(x)$  accounts for the presence of a hard wall at $x=0$. In the steady state, the probability distributions in the $u = \pm 1$ states satisfy
\begin{eqnarray}
        & D \partial_x^2 P_+(x) -v \partial_x P_+(x) + \frac{1}{\tau}\left(P_-(x) - P_+(x)\right) = 0 \,, \nonumber \\ 
        & D \partial_x^2 P_-(x) + v \partial_x P_-(x) + \frac{1}{\tau}\left(P_+(x) - P_-(x)\right) = 0 \,,
\end{eqnarray}
together with the no-flux boundary conditions at the wall
\begin{eqnarray}
        & D \partial_x P_+(0) -v P_+(0) = 0 \,, \nonumber \\ 
        & D \partial_x P_-(0) + v P_-(0)  = 0 \,,
\end{eqnarray}
and the boundary conditions at infinity
\begin{eqnarray}
    P_+(\infty) = P_-(\infty) = \frac{1}{2} \,.
\end{eqnarray}
Let $\phi(x) = P_+(x) + P_-(x)$ be the total density and $Q(x) = P_+(x) - P_-(x)$ the polarization. We get
\begin{eqnarray}
        & D \partial_x^2 \phi(x) - v \partial_x Q(x) = 0 \,, \nonumber \\ 
        & D \partial_x^2 Q(x) - v \partial_x \phi(x) + \frac{2}{\tau}Q(x) = 0 \,,
\end{eqnarray}
together with 
\begin{eqnarray}\label{eq:boundary_condition}
        & D \partial_x \phi(0) - v Q(0) = 0 \,, \nonumber \\ 
        & D \partial_x Q(0) - v \phi(0)  = 0 \,.
\end{eqnarray}
Therefore, 
\begin{eqnarray}\label{eq:polarization_bc}
    Q(x) = \frac{D}{v}\partial_x \phi(x) \,,
\end{eqnarray}
so that
\begin{eqnarray}
    \partial_x^3 \phi(x) - \left(\frac{v^2}{D^2} + \frac{2}{D\tau}\right)\partial_x \phi(x) = 0 \,.
\end{eqnarray}
We deduce from the above equation that the density profile decays exponentially with the distance from the wall, 
\begin{eqnarray}
    \phi(x) = 1 + c \exp\left(-\sqrt{\frac{v^2}{D^2} + \frac{2}{D\tau}}\,x\right) \,,
\end{eqnarray}
where we used the boundary condition $\phi(\infty) = 1$ and where $c$ is an integration constant that remains to be determined. To proceed, we use the no-flux boundary condition in the second line of Eq.~\eqref{eq:boundary_condition} together with Eq.~\eqref{eq:polarization_bc} to get
\begin{eqnarray}
    \partial^2_x \phi(0) - \frac{v^2}{D^2} \phi(0) = 0 \,.
\end{eqnarray}
This allows us to determine $c$ and to get
\begin{eqnarray}
    \phi(x) = 1 + \frac{\tau  v^2}{2 D} \exp\left(-\sqrt{\frac{v^2}{D^2} + \frac{2}{D\tau}}\,x\right) \,.
\end{eqnarray}

\bibliography{biblio_confinement}